\documentclass[10pt,aps,prc,twocolumn,superscriptaddress,floatfix,nofootinbib]{revtex4-2}
\usepackage{braket}
\usepackage{amsthm}
\usepackage{ascmac}
\usepackage{amssymb, amsmath}
\usepackage{bm}
\usepackage{physics}
\usepackage{autobreak}

\usepackage{graphicx}
\usepackage{times}
\usepackage[T1]{fontenc}
\usepackage{ulem}
\usepackage{color}
\usepackage{multirow}
\usepackage{geometry}
\newcommand{\up}{\!\uparrow}
\newcommand{\down}{\!\downarrow}

\begin{document}

\title{Superfluid extension of the self-consistent time-dependent band theory for neutron star matter:\\
Anti-entrainment versus superfluid effects in the slab phase}

\author{Kenta Yoshimura}
\email[]{yoshimura.k.ak@m.titech.ac.jp}
\affiliation{Department of Physics, School of Science, Tokyo Institute of Technology, Tokyo 152-8551, Japan}

\author{Kazuyuki Sekizawa}
\email[]{sekizawa@phys.titech.ac.jp}
\affiliation{Department of Physics, School of Science, Tokyo Institute of Technology, Tokyo 152-8551, Japan}
\affiliation{Nuclear Physics Division, Center for Computational Sciences, University of Tsukuba, Ibaraki 305-8577, Japan}
\affiliation{RIKEN Nishina Center, Saitama 351-0198, Japan}

\date{June 19, 2024}

\begin{abstract}
\begin{description}
\item[Background]
The inner crust of neutron stars consists of a Coulomb lattice of neutron-rich nuclei,
immersed in a sea of superfluid neutrons with background relativistic electron gas.
A proper quantum mechanical treatment for such a system under a periodic potential
is the band theory of solids. The effect of band structure on the effective mass
of dripped neutrons, the so-called \textit{entrainment effect}, is currently in a debatable
situation, and it has been highly desired to develop a microscopic nuclear band theory taking
into account neutron superfluidity in a fully self-consistent manner.

\item[Purpose]
The main purpose of the present work is twofold: 1) to develop a formalism of the
time-dependent self-consistent band theory, taking fully into account nuclear superfluidity,
based on time-dependent density functional theory (TDDFT) extended for superfluid systems,
and 2) to quantify the effects of band structure and superfluidity on the entrainment phenomenon,
applying the formalism to the slab phase of the inner crust of neutron stars.

\item[Methods]
The fully self-consistent time-dependent band theory, proposed in a previous work
[K.~Sekizawa, S.~Kobayashi, and M.~Matsuo, Phys. Rev. C \textbf{105}, 045807 (2022)],
is extended for superfluid systems. To this end, a superfluid TDDFT with a local treatment
of pairing, known as time-dependent superfluid local density approximation (TDSLDA),
is formulated in the coordinate space with a Skyrme-type energy density functional,
adopting the Bloch's boundary condition. A real-time method is employed to extract the
collective masses of a slab and of protons, which in turn quantify the conduction
neutron number density and the neutron effective mass, i.e., the entrainment effect.

\item[Results]
Static calculations have been performed for a range of baryon number density
($n_\text{b}=0.04$--0.07\,fm$^{-3}$) under the $\beta$-equilibrium condition with and
without superfluidity, for various inter-slab spacings. From the results, we find that
the system gains energy through the formation of Cooper pairs for all densities examined,
which supports the existence of superfluidity in the inner crust of cold neutron stars.
From a response of the system to an external potential, we dynamically extract the collective
masses of a slab and of protons immersed in neutron superfluid. The obtained results
show the collective mass of a slab is substantially reduced by 57.5–82.5\% for
$n_\text{b}=0.04$--0.07\,fm$^{-3}$, which corresponds to an enhancement of conduction
neutron number density and, thus, to a \textit{reduction} of the neutron effective mass,
which we call the \textit{anti-entrainment} effect. A comparison of the results with and
without superfluidity reveals that superfluidity slightly enhances the anti-entrainment
effects for the slab phase of neutron-star matter. We discuss a novel phenomenon associated
with superfluidity, that is, quasiparticle resonances in the inner crust, which are absent in
normal systems.

\item[Conclusions]
Our fully self-consistent, microscopic, superfluid band theory calculations based
on (TD)DFT showed that the effective mass of dripped neutrons is \textit{reduced}
by about 20--40\% for $n_\text{b}=0.04$--0.07\,fm$^{-3}$ because of the band structure
effects, and superfluidity slightly \textit{enhances} the reduction.

\end{description}
\end{abstract}


\maketitle

\section{INTRODUCTION}
The band theory of solids, with the aid of Kohn-Sham density functional theory
(DFT) \cite{DFT1,DFT2,DFT3}, has become a standard tool in material science to
understand and predict abundant properties of a wide variety of matters.
Its time-dependent extension based on time-dependent DFT (TDDFT) \cite{DFT4,DFT5,DFT6}
has also been extensively developed to explore electron dynamics under a strong
laser field (see, e.g., Refs.~\cite{Yabana(1996),Bertsch(2000),Yabana(2012),
Yamada(2019),Octopus,SALMON} and references therein). Recently, the band theory
of solids has been applied for studying properties of the inner crust of neutron
stars \cite{Carter(2005),Chamel(2005),Chamel(2006),Chamel(2007),Chamel(2012),
Kashiwaba(2019),Sekizawa(2022)}, where nuclear bundles form crystalline structures,
just in a similar way as terrestrial solids, which coexists with neutron superfluid
and relativistic electron gas. Along the line with the previous work \cite{Sekizawa(2022)},
this paper aims to develop a fully self-consistent, microscopic framework to describe
structure and dynamics of the inner crust of neutron stars, taking fully into
account both the band structure and superfluid effects on the same footing.

\begin{table*}[t]
  \centering
  \caption{
  A summary of the complicated situation concerning the band structure effects
  in the inner crust of neutron stars. This work, with future extensions to 2D
  and 3D systems, aims to provide conclusive values of the neutron effective mass
  throughout the inner crust of neutron stars.
  }\vspace{3mm}
  \begin{tabular*}{\textwidth}{@{\extracolsep{\fill}}ccccccc}
    \hline\hline
    Author(s) (Year)                                  & Dimension & Self-consistency & Superfluidity & $m_n^\star/m_n$ & $n_\text{b}$ (fm$^{-3}$) & Ref.      \\
    \hline
    \multirow{2}{*}{Carter, Chamel \& Haensel (2005)} & 1D     & --         & --         & 1.02--1.03   & 0.074--0.079      & \multirow{2}{*}{\cite{Carter(2005)}} \\
                                                      & 2D     & --         & --         & 1.11--1.40   & 0.058--0.072      &                                      \\
    Chamel    (2005)                                  & 3D     & --         & --         & 1.07--15.4   & 0.03--0.086       & \cite{Chamel(2005)}                  \\
    Chamel    (2012)                                  & 3D     & --         & --         & 1.21--13.6   & 0.0003--0.08      & \cite{Chamel(2012)}                  \\
    Kashiwaba \& Nakatsukasa (2019)                   & 1D     & \checkmark & --         & 0.65--0.75   & 0.07--0.08$^\dag$ & \cite{Kashiwaba(2019)}               \\
    Sekizawa, Kobayashi \& Matsuo (2022)              & 1D     & \checkmark & --         & 0.59         & 0.04$^\ddag$      & \cite{Sekizawa(2022)}                \\
    Yoshimura \& Sekizawa (2024)                      & 1D     & \checkmark & \checkmark & 0.58         & 0.07$^\dag$       & This work                            \\
                                                      & 2D, 3D & \checkmark & \checkmark & ??           & $\lesssim0.07$    & Future works                         \\
    \hline\hline\\[-4mm]
    \multicolumn{6}{l}{\footnotesize $\dag $\,Where appearance of the slab phase is expected.}\\
    \multicolumn{6}{l}{\footnotesize $\ddag$\,With a fixed proton fraction, $Y_p=0.1$.}\\
  \end{tabular*}
  \label{Table:situation}
\end{table*}

The band theory of solids may not yet be a popular approach in the context of
neutron-star studies. While a first indication of necessity of such a calculation
was made in 1994 \cite{Oyamatsu(1994)}, realistic band theory calculations were achieved
in 2005 for slab and rod phases \cite{Carter(2005)}, followed by three-dimensional
(3D) calculations for Coulomb lattices of spherical nuclei \cite{Chamel(2005),Chamel(2012)}.
It has been shown that, based on band theory calculations on top of a nuclear
potential obtained with the Thomas-Fermi-type approximation \cite{Onsi(2008)},
the effective mass of dripped neutrons is increased by factors of 1.02--1.03 and
1.11--1.40 in the slab and rod phases, respectively. Furthermore, the 3D calculations
showed that band structure effects always \textit{enhance} the neutron effective
mass and most strikingly in a certain low-density region (0.02\,fm$^{-3}\lesssim
$\,$n_\text{b}$\,$\lesssim$\,0.04\,fm$^{-3}$) it was found to be nearly 10 times
or more larger than the bare neutron mass \cite{Chamel(2012)}. The increase of
neutron effective mass is associated with the Brag scattering of dripped neutrons off the periodic
potential, which immobilize those otherwise-free neutrons. The latter effect is
called the \textit{entrainment effect}. The notable change of the neutron effective
mass turned out to affect various interpretations of astrophysical phenomena of
neutron stars such as pulsar glitches \cite{Andersson(2012),Chamel(2013)glitch,
Haskell(2015)} and thermal as well as crustal properties \cite{Chamel(2009),
Chamel(2013),Kobyakov(2013),Kobyakov(2016),Durel(2018)}, and it has attracted
increasing interests over the years. (See, Ref.~\cite{Chamel(2017)}, for a review
of band calculations of Chamel et al.\ and discussion on related topics.)

The situation regarding the band structure effects on dripped neutrons in
the inner crust of neutron stars is still highly controversial (see Table~\ref{Table:situation}).
While the band theory calculations assume a perfect crystalline structure, disorder of crystal
may reduce the band structure effects \cite{Sauls(2020)}. Apart from the possible
disorder effects, recently, fully self-consistent, microscopic band theory calculations
based on DFT and its time-dependent extension (TDDFT) have been achieved for
the slab phase of nuclear matter \cite{Kashiwaba(2019),Sekizawa(2022)}. In
Refs.~\cite{Kashiwaba(2019),Sekizawa(2022)}, based on both static and dynamic
calculations, respectively, the neutron effective mass was found to be \textit{reduced}
for the slab phase, which is called the \textit{anti-entrainment} effects
\cite{Sekizawa(2022)}. The latter observation is contradictory to the results
of Ref.~\cite{Carter(2005)}, which may be partly due to an improper definition
of ``free'' neutron density for the slab and rod phases in the work of Carter
et al.\ \cite{Carter(2005)}, as pointed out in Ref.~\cite{Kashiwaba(2019)}.
Further applications of the fully self-consistent band theory to higher
spatial dimensions have been highly desired.

However, in the aforementioned works, pairing correlations were neglected and
it is not at all obvious how superfluidity affects the entrainment effect.
In Ref.~\cite{Carter(2005)BCS}, it was argued, within the Bardeen-Cooper-Schrieffer
(BCS) approximation, that pairing correlations do not affect much the results and
the strong entrainment would remain. In later studies, pairing effects were studied
for a 1D periodic potential within the BCS as well as Hartree-Fock-Bogoliubov (HFB)
[also called Bogoliubov-de Genns (BdG)] approximations, showing that the BCS treatment
is not enough and a self-consistent treatment of pairing correlations is essentially
important to correctly quantify the entrainment effect \cite{Watanabe(2017),Minami(2022)}.
However, none of those studies are fully self-consistent. In neutron-star matter,
there is no ``external'' potential (except the gravitational one that is negligible
in investigating nuclear-scale microphysics) and neutrons and protons are self-organizing
to arrange a variety of crystalline structures. Therefore, it is an imperative task
to develop a fully self-consistent microscopic framework that includes superfluidity
to draw a clear conclusion on the magnitude of the entrainment effect.

In the present paper, before extending the framework of Ref.~\cite{Sekizawa(2022)}
to higher spatial dimensions, we shall first develop a formalism of fully self-consistent,
microscopic time-dependent superfluid band calculations based on TDDFT for superfluid systems.
According to the Hohenberg-Kohn theorem \cite{DFT1} with the Kohn-Sham scheme \cite{DFT2},
DFT can, in principle, be an exact approach to quantum many-body problems. Its time-dependent
extension \cite{DFT5,DFT6}, based on the Runge-Gross theorem \cite{DFT4}, allows us to describe
complex non-linear dynamics as well as excited states. While (TD)DFT for superfluid
(superconducting) systems was initially formulated with a non-local pair potential
$\Delta(\bm{r},\bm{r}')$ \cite{Oliveira(1988),Wacker(1994)}, subsequent developments
of its local treatment with a proper pairing renormalization scheme resulted in the
so-called (time-dependent) superfluid local density approximation [(TD)SLDA]
\cite{Kurth(1999),Bulgac(2002)1,Bulgac(2002)2,Bulgac(2009),Bulgac(2011)}. In the
nuclear physics context, on the other side, (TD)DFT was originally developed as
mean-field theories with effective nucleon-nucleon interactions, like the Skyrme
(TD)HF approach \cite{Bender(2003),Simenel(review),Nakatsukasa(review),TDHF-review(2018),
Stevenson(2019),Sekizawa(2019)}. Because of the historical reason, one may
confuse (TD)SLDA as (TD)HFB with a zero-range effective pairing interaction.
We note, however, that the pairing renormalization scheme \cite{Bulgac(2002)1,
Bulgac(2002)2} makes the theory cutoff independent, if it is taken to be sufficiently
large, and allows one to work with a local pairing field $\Delta(\bm{r})$ within a
well-defined theoretical framework. Thus, in the same way as one regards Skyrme (TD)HF
as (TD)DFT, we can regard Skyrme (TD)HFB with a zero-range pairing interaction as
superfluid (TD)DFT, if such a proper pairing renormalization scheme is adopted.
TDSLDA \cite{Bulgac(2012),Bulgac(2019)} has been successfully applied not only
for nuclear systems \cite{Bulgac(2016),VN-int(2016),MSW(2017),Bulgac(2019),
Magierski(2022)}, but also for cold-atomic systems \cite{Bulgac(2014),
Wlazlowski(2015),Wlazlowski(2018),Hossain(2022),Tuzemen(2023),Barresi(2023)}.
In this work, we develop a fully self-consistent (time-dependent) superfluid
band theory based on (TD)SLDA, imposing the Bloch's boundary condition to
quasiparticle wave functions, which corresponds to an extension of the previous
work \cite{Sekizawa(2022)} for superfluid systems. By applying the formalism
to the slab phase of neutron-star matter under the $\beta$-equilibrium condition,
we demonstrate the validity of our formalism and shed new light on the role of
superfluidity in the entrainment phenomenon.

The article is organized as follows.
In Sec.~\ref{sec:formulation}, we explain detailed formalisms of the self-consistent
time-dependent superfluid band theory for the inner crust of neutron stars, and
provide computational details in Sec.~\ref{Sec:ComptDetails}.
In Sec.~\ref{Sec:Results}, we present the results of band theory calculations
for the slab phase of nuclear matter under the $\beta$-equilibrium condition.
In Sec.~\ref{Sec:Conclusion}, conclusions and prospect are given.

\section{Formulation}\label{sec:formulation}

\subsection{The HFB theory}

\subsubsection{The matrix representation}

Let us first succinctly recapitulate the general framework of the HFB theory,
clarifying our notations. Here we start with a generic Hamiltonian in the
second quantization form,
\begin{equation}
\hat{H} = \sum_{kl}t_{kl}\hat{a}_k^\dag\hat{a}_l
+ \frac{1}{4}\sum_{klmn}\bar{v}_{klmn}\hat{a}_k^\dag\hat{a}_l^\dag\hat{a}_n\hat{a}_m,
\end{equation}
where $\hat{a}_k$ and $\hat{a}_k^\dag$ are particle annihilation and
creation operators, respectively, that obey the Fermionic anticommutation
relations: i.e., $\{\hat{a}_k,\hat{a}_l^\dag\}=\delta_{kl}$ and $\{\hat{a}_k,
\hat{a}_l\}=\{\hat{a}_k^\dag,\hat{a}_l^\dag\}=0$. $t_{kl}$ and $\bar{v}_{klmn}$
$(\equiv v_{klmn}-v_{klnm})$ are usual matrix elements of a one-body kinetic
energy operator and of a two-body interaction, respectively, where the latter one
is antisymmetrized for convenience. In the HFB theory, quasiparticle annihilation
and creation operators, $\hat{\beta}_\mu$ and $\hat{\beta}_\mu^\dag$, respectively,
are introduced via the Bogoliubov transformation of $\hat{a}_k$ and $\hat{a}_k^\dag$:
\begin{eqnarray}
\mqty(\hat{\boldsymbol{\beta}}\\ \hat{\boldsymbol{\beta}}^\dag)
= \mathcal{W}^\dag \mqty(\hat{\boldsymbol{a}}\\ \hat{\boldsymbol{a}}^\dag).\label{Eq:BogoliubovTF}
\end{eqnarray}
Here we have introduced column vectors,
$\hat{\boldsymbol{\beta}}\equiv\bigl(\hat{\beta}_1, \hat{\beta}_2, \dots, \hat{\beta}_M\bigr)^\mathrm{T}$,
$\hat{\boldsymbol{\beta}}^\dag\equiv\bigl(\hat{\beta}_1^\dag, \hat{\beta}_2^\dag,\dots, \hat{\beta}_M^\dag\bigr)^\mathrm{T}$,
$\hat{\boldsymbol{a}}\equiv\bigl(\hat{a}_1, \hat{a}_2, \dots, \hat{a}_M\bigr)^\mathrm{T}$, and
$\hat{\boldsymbol{a}}^\dag\equiv\bigl(\hat{a}_1^\dag, \hat{a}_2^\dag,\dots, \hat{a}_M^\dag\bigr)^\mathrm{T}$,
to simplify the notation, where $M$ corresponds to the dimension of basis states. The $2M\times2M$
Bogoliubov transformation matrix $\mathcal{W}$ can be written as
\begin{equation}
\mathcal{W} = \mqty(U& V^*\\ V& U^*),
\end{equation}
where $U$ and $V$ are $M\times M$ matrices. The Bogoliubov transformation
matrix is unitary, i.e.\ $\mathcal{W^\dag W}=\mathcal{WW^\dag}=I_{2M}$, with $I_{2M}$
being a $2M$-dimensional identity matrix. The latter property ensures that
$\hat{\beta}_\mu$ and $\hat{\beta}_\mu^\dag$ also obey the Fermionic anticommutation
relations. One can write down explicitly the quasiparticle annihilation and
creation operators, respectively, as follows:
\begin{eqnarray}
\hat{\beta}_\mu &=& \sum_i\Bigl(
U^*_{i\mu}\,\hat{a}_i + V^*_{i\mu}\,\hat{a}_i^\dagger \Bigr),\label{Eq:def_beta}\\
\hat{\beta}_\mu^\dagger &=& \sum_i\Bigl(
U_{i\mu}\,\hat{a}_i^\dagger + V_{i\mu}\,\hat{a}_i\Bigr).\label{Eq:def_beta_dagger}
\end{eqnarray}
Note that we use Greek indices (such as $\mu,\nu,\dots$) for labeling
positive-energy quasiparticle states, while Roman indices (such as $i,j,\dots$)
are used for labeling single-particle states, except some obvious cases.

The HFB state, the trial many-body wave function for a variation, is defined as
a vacuum of quasiparticles, i.e.\ $\hat{\beta}_\mu\big|\text{HFB}\bigr>=0$ for
all $\mu$. Based on the variational principle, one can derive the well-known
HFB equation:
\begin{equation}
\mqty(h-\lambda I_M & \Delta \\ -\Delta^* & -h^*+\lambda I_M)\mqty(U_\mu \\ V_\mu)
= E_\mu \mqty(U_\mu \\ V_\mu),
\label{Eq:BdG}
\end{equation}
where $h=t+\Gamma$ denotes the single-particle Hamiltonian matrix with
a matrix for the mean-field potential $\Gamma$, $\Delta$ is a matrix for
the pair potential, $\lambda$ is the chemical potential, and $U_\mu$
and $V_\mu$ represent the $\mu$-th column of the $U$ and $V$ matrices,
respectively. The $(k,l)$ component of the mean-field potential and the
pair potential matrices are defined, respectively, as follows:
\begin{eqnarray}
\Gamma_{kl} &=& \sum_{mn}\bar{v}_{kmln}\rho_{nm},\\
\Delta_{kl} &=& \frac{1}{2}\sum_{mn}\bar{v}_{klmn}\kappa_{mn}.
\end{eqnarray}
Here the one-body density matrix, $\rho$, and the pairing tensor, $\kappa$,
are, respectively, given by
\begin{eqnarray}
\rho_{kl} &=& \bigl<\hat{a}_l^\dag\hat{a}_k\bigr> = (V^*V^\mathrm{T})_{kl},\\[2mm]
\kappa_{kl} &=& \bigl<\hat{a}_l\hat{a}_k\bigr> = (V^*U^\mathrm{T})_{kl},
\end{eqnarray}
where the brackets, $\bigl<\cdots\bigr>$, represent an expectation value in
the HFB state. This is the usual matrix representation of the HFB theory.
By definition, $\rho$ is hermitian ($\rho^\dag=\rho$) and $\kappa$ is
skew symmetric ($\kappa^\mathrm{T}=-\kappa$).

\subsubsection{Treatment of a system with certain symmetries in
the coordinate-space representation}\label{Sec:SLDAsymm}

Next, let us consider a case where a system possesses certain symmetries,
and introduce the coordinate-space representation of the HFB theory. The
formulas given here are actually useful to formulate the superfluid band
theory with the Bloch's boundary condition in Sec.~\ref{Sec:SuperfluidBandTheory}.
When a system possesses symmetries, the Hamiltonian commutes with operators
associated with the corresponding symmetric transformations and there are
conserved quantities with which one can classify quantum states. For instance,
for a system with the spherical symmetry, the orbital angular momentum $L$
and its projection $m$ are conserved, being good quantum numbers. In such
a case, states with different values of $(L,m)$ are not mixed, and the
Hamiltonian matrix can be arranged to have a block diagonal form. In the
following we shall denote such a set of arbitrary good quantum numbers as
$\Omega$.

Since single- and quasi-particle states can be classified according to
the set of quantum numbers, $\Omega$, we may explicitly indicate it as
$\mu\rightarrow\{\nu\,\Omega\}$ and $i\rightarrow\{j\,\Omega\}$. For a
system with symmetries, the quasiparticle annihilation and creation operators,
Eqs.~\eqref{Eq:def_beta} and \eqref{Eq:def_beta_dagger}, respectively,
can be written as
\begin{eqnarray}
\hat{\beta}_{\nu\Omega} &=& \sum_j\Bigl(
U^*_{j\Omega,\nu\Omega}\,\hat{a}_{j\Omega} + V^*_{j\bar{\Omega},\nu\Omega}\,\hat{a}_{j\bar{\Omega}}^\dagger \Bigr),\label{Eq:beta_nuOmega}\\
\hat{\beta}_{\nu\Omega}^\dagger &=& \sum_j\Bigl(
U_{j\Omega,\nu\Omega}\,\hat{a}_{j\Omega}^\dagger + V_{j\bar{\Omega},\nu\Omega}\,\hat{a}_{j\bar{\Omega}}\Bigr).\label{Eq:beta_daggar_nuOmega}
\end{eqnarray}
Note that $\bar{\Omega}$, which appeared as a subscript of the second term in
the parentheses in Eqs.~\eqref{Eq:beta_nuOmega} and \eqref{Eq:beta_daggar_nuOmega},
stands for the same set of quantum numbers as $\Omega$, but any ``countable''
quantum numbers involved in it have opposite sign [e.g., for the case of
$\Omega=(L,m)$, $\bar{\Omega}=(L,-m)$]. It is simply because a ``hole'' of
a state with countable quantum number(s) could be characterized like a particle
which has the opposite sign for the countable quantum number(s). In this way,
the quantum numbers in the left- and right-hand-side of Eqs.~\eqref{Eq:beta_nuOmega}
and \eqref{Eq:beta_daggar_nuOmega} are being consistent. It should be noted that
Eqs.~\eqref{Eq:beta_nuOmega} and \eqref{Eq:beta_daggar_nuOmega} mean the original
$U$, $V$ matrices are now in a block diagonal form. That is, $U$ relates single-
and quasi-particle states with the same $\Omega$, while $V$ relates those with
$\Omega$ and $\bar{\Omega}$, being $U_{j\Omega,\nu\Omega'}=U_{j\Omega,\nu\Omega}\,\delta_{\Omega\Omega'}$
and $V_{j\Omega,\nu\Omega'}=V_{j\Omega,\nu\bar{\Omega}}\,\delta_{\bar{\Omega}\Omega'}$.

To obtain the coordinate-space representation with the spin degree of freedom,
let us introduce the field operators,
\begin{eqnarray}
\hat{\psi}(\bm{r}\sigma) &=& \sum_{j\Omega}\phi_{j\Omega}(\bm{r}\sigma)\,\hat{a}_{j\Omega},\label{Eq:hat_psi}\\
\hat{\psi}^\dag(\bm{r}\sigma) &=& \sum_{j\Omega}\phi_{j\Omega}^*(\bm{r}\sigma)\,\hat{a}_{j\Omega}^\dag.\label{Eq:hat_psi_dagger}
\end{eqnarray}
Note that the summation is taken over all single-particle states ($i=\{j\,\Omega\}$),
by definition. Here $\phi_{j\Omega}(\bm{r}\sigma)\equiv\bigl<\bm{r}\sigma\big|j\Omega\bigr>
=\bigl<\bm{r}\sigma\big|\hat{a}_{j\Omega}^\dag\big|0\bigr>$ denotes the
single-particle wave function, where $\big|0\bigr>$ is the vacuum state.
The field operator creates a particle with spin $\sigma$ ($=$\,$\uparrow$ or
$\down$) at a position $\bm{r}$ in the vacuum, i.e.\ $\big|\bm{r}\sigma\bigr>
=\hat{\psi}^\dag(\bm{r}\sigma)\big|0\bigr>$. Using the orthonormal properties
of the single-particle wave functions, $\bigl<\phi_k\big|\phi_l\bigr>\equiv
\sum_\sigma\int \phi_k^*(\bm{r}\sigma)\phi_l(\bm{r}\sigma)\dd\bm{r}=\delta_{kl}$,
with Eqs.~\eqref{Eq:hat_psi} and \eqref{Eq:hat_psi_dagger}, the particle annihilation
and creation operators can be represented, respectively, as follows:
\begin{eqnarray}
\hat{a}_{j\Omega} &=& \sum_\sigma\int \phi_{j\Omega}^*(\bm{r}\sigma)\hat{\psi}(\bm{r}\sigma)\,\dd\bm{r},\label{Eq:a_jOmega}\\
\hat{a}_{j\Omega}^\dagger &=& \sum_\sigma\int \phi_{j\Omega}(\bm{r}\sigma)\hat{\psi}^\dag(\bm{r}\sigma)\,\dd\bm{r}.\label{Eq:a_daggar_jOmega}
\end{eqnarray}
Substituting Eqs.~\eqref{Eq:a_jOmega} and \eqref{Eq:a_daggar_jOmega} into
Eqs.~\eqref{Eq:beta_nuOmega} and \eqref{Eq:beta_daggar_nuOmega}, one finds
\begin{eqnarray}
\hat{\beta}_{\nu\Omega} &=& \sum_\sigma\int\Bigl(
u^*_{\nu\Omega}(\bm{r}\sigma)\hat{\psi}(\bm{r}\sigma) + v^*_{\nu\Omega}(\bm{r}\sigma)\hat{\psi}^\dag(\bm{r}\sigma) \Bigr)\dd\bm{r},\;\;\;\;\;\;\;\\
\hat{\beta}_{\nu\Omega}^\dagger &=& \sum_\sigma\int\Bigl(
u_{\nu\Omega}(\bm{r}\sigma)\hat{\psi}^\dagger(\bm{r}\sigma) + v_{\nu\Omega}(\bm{r}\sigma)\hat{\psi}(\bm{r}\sigma)\Bigr)\dd\bm{r},
\end{eqnarray}
where
\begin{eqnarray}
u_{\nu\Omega}(\bm{r}\sigma) &\equiv& \sum_j U_{j\Omega,\nu\Omega}\,\phi_{j\Omega}(\bm{r}\sigma),
\label{Eq:def_qpwf_u}\\
v_{\nu\Omega}(\bm{r}\sigma) &\equiv& \sum_j V_{j\bar{\Omega},\nu\Omega}\,\phi^*_{j\bar{\Omega}}(\bm{r}\sigma).
\label{Eq:def_qpwf_v}
\end{eqnarray}
Those are the coordinate-space representation of the $u$ and $v$
components of the quasiparticle wave functions. The quasiparticle
wave functions are normalized to be
\begin{equation}
\sum_\sigma\int \Bigl[|u_{\nu\Omega}(\bm{r}\sigma)|^2
+ |v_{\nu\Omega}(\bm{r}\sigma)|^2 \Bigr]\,\dd\bm{r} = 1.
\label{Eq:normalization_qpwf}
\end{equation}
In the same way as in the matrix representation, a variational calculation
leads to the corresponding coordinate-space representation of the HFB equation:
\begin{widetext}
\begin{eqnarray}
\int \!d\bm{r}'\!
\begin{pmatrix}
\hat{h}_{\uparrow\up}(\bm{r},\bm{r}')-\lambda\delta_{\bm{r},\bm{r}'} & \hat{h}_{\uparrow\down}(\bm{r},\bm{r}') & 0 & \Delta(\bm{r},\bm{r}')\\
\hat{h}_{\downarrow\up}(\bm{r},\bm{r}') & \hat{h}_{\downarrow\down}(\bm{r},\bm{r}')-\lambda\delta_{\bm{r},\bm{r}'} & -\Delta(\bm{r},\bm{r}') & 0\\
0 & -\Delta^*(\bm{r},\bm{r}') & -\hat{h}^*_{\uparrow\up}(\bm{r},\bm{r}')+\lambda\delta_{\bm{r},\bm{r}'} & -\hat{h}^*_{\uparrow\down}(\bm{r},\bm{r}')\\
\Delta^*(\bm{r},\bm{r}') & 0 & -\hat{h}^*_{\downarrow\up}(\bm{r},\bm{r}') & -\hat{h}^*_{\downarrow\down}(\bm{r},\bm{r}')+\lambda\delta_{\bm{r},\bm{r}'}
\end{pmatrix}\!
\begin{pmatrix}
u_{\nu\Omega}(\bm{r}'\up)\\
u_{\nu\Omega}(\bm{r}'\down)\\
v_{\nu\Omega}(\bm{r}'\up)\\
v_{\nu\Omega}(\bm{r}'\down)
\end{pmatrix}\!
= E_{\nu\Omega}\!
\begin{pmatrix}
u_{\nu\Omega}(\bm{r}\up)\\
u_{\nu\Omega}(\bm{r}\down)\\
v_{\nu\Omega}(\bm{r}\up)\\
v_{\nu\Omega}(\bm{r}\down)
\end{pmatrix},
\label{Eq:HFB_coordinate-space}
\end{eqnarray}
\end{widetext}
where $\delta_{\bm{r},\bm{r}'}\equiv\delta(\bm{r}-\bm{r}')$. In the
coordinate-space representation, the number and anomalous densities
are defined as $n(\bm{r})\equiv\sum_\sigma n(\bm{r}\sigma,\bm{r}\sigma)$
and $\kappa(\bm{r},\bm{r}')\equiv\kappa(\bm{r}\up,\bm{r}'\down)$,
respectively, where
\begin{eqnarray}
n(\bm{r}\sigma,\bm{r}'\sigma')
&\equiv& \bigl<\hat{\psi}^\dag(\bm{r}'\sigma')\hat{\psi}(\bm{r}\sigma)\bigr> \nonumber\\[1.5mm]
&=& \sum_{i\Omega_1j\Omega_2\nu\Omega}\phi^*_{i\Omega_1}(\bm{r}'\sigma')\phi_{j\Omega_2}(\bm{r}\sigma) V^*_{j\Omega_2,\nu\Omega}V^\mathrm{T}_{\nu\Omega,i\Omega_1} \nonumber\\
&=& \sum_{\nu\Omega} v^*_{\nu\Omega}(\bm{r}\sigma)v_{\nu\Omega}(\bm{r}'\sigma'),
\label{Eq:def_n}
\end{eqnarray}
\begin{eqnarray}
\kappa(\bm{r}\sigma,\bm{r}'\sigma')
&\equiv& \bigl<\hat{\psi}(\bm{r}'\sigma')\hat{\psi}(\bm{r}\sigma)\bigr> \nonumber\\[1.5mm]
&=& \sum_{i\Omega_1j\Omega_2\nu\Omega}\phi_{i\Omega_1}(\bm{r}'\sigma')\phi_{j\Omega_2}(\bm{r}\sigma) V^*_{j\Omega_2,\nu\Omega}U^\mathrm{T}_{\nu\Omega,i\Omega_1} \nonumber\\
&=& \sum_{\nu\Omega} v^*_{\nu\Omega}(\bm{r}\sigma)u_{\nu\Omega}(\bm{r}'\sigma').
\label{Eq:def_kappa}
\end{eqnarray}

When the single-particle Hamiltonian contains no terms that mix spin
states (i.e.\ $\hat{h}_{\up\down}=\hat{h}_{\down\up}=0$),
the HFB equation \eqref{Eq:HFB_coordinate-space} can be decomposed
into two equations with a half of the original dimension as
\begin{eqnarray}
\int\! d\bm{r}'&&\hspace{-4.5mm}
\begin{pmatrix}
\hat{h}(\bm{r},\bm{r}')-\lambda\delta_{\bm{r},\bm{r}'}\hspace{-3mm} & \Delta(\bm{r},\bm{r}')\\
\Delta^*(\bm{r},\bm{r}') & -\hat{h}^*(\bm{r},\bm{r}')+\lambda\delta_{\bm{r},\bm{r}'}
\end{pmatrix}
\begin{pmatrix}
u_{\nu\Omega}(\bm{r}'\up)\\
v_{\nu\Omega}(\bm{r}'\down)
\end{pmatrix} \nonumber\\[1mm]
&&\hspace{34mm}= E_{\nu\Omega}
\begin{pmatrix}
u_{\nu\Omega}(\bm{r}\up)\\
v_{\nu\Omega}(\bm{r}\down)
\end{pmatrix},
\label{Eq:HFB_coordinate-space_reduced}
\end{eqnarray}
where $\hat{h}=\hat{h}_{\uparrow\up}=\hat{h}_{\downarrow\down}$. From a
diagonalization of the HFB matrix, one obtains not only the quasiparticle
states, but also the quasihole states with negative eigenvalues, $-E_{\nu\Omega}$.
Thanks to this so-called quasiparticle-quasihole symmetry, the other spin
component of quasiparticle wave functions, i.e.\ $\bigl(v_{\nu\Omega}^*(\bm{r}\up),
u_{\nu\Omega}^*(\bm{r}\down)\bigr)^\mathrm{T}$ for $-E_{\nu\Omega}$,
can be found in the negative energy states.

\subsection{Superfluid band theory with a Skyrme-type EDF}\label{Sec:SuperfluidBandTheory}

\subsubsection{The Bloch's boundary condition}\label{Sec:SLDABloch}

In this section, we formulate the band theory of solids for superfluid
systems. What one has to do is to combine the Bloch's boundary condition
with the HFB framework. The essence of the band theory is to impose the
periodicity of the crystal to the wave functions of the system. According
to the Floquet-Bloch theorem, it can be achieved, representing the
single-particle wave functions by modulated plane waves \cite{Ashcroft-Mermin},
\begin{equation}
\phi_{j\bm{k}}^{(q)}(\bm{r}\sigma)
= \frac{1}{\sqrt{\mathcal{V}}}\widetilde{\phi}_{j\bm{k}}^{(q)}(\bm{r}\sigma)e^{\mathrm{i}\bm{k\cdot r}},
\label{Eq:spwf_Bloch_k}
\end{equation}
where $\mathcal{V}$ stands for the volume of a unit cell and $\bm{k}$
is the Bloch wave vector. In this section, we explicitly indicate
the isospin degree of freedom by an index $q$, where $q=n$ for neutrons
and $q=p$ for protons. The periodicity of the system is then
encoded into the function $\widetilde{\phi}_{j\bm{k}}^{(q)}(\bm{r}\sigma)$ as
\begin{equation}
\widetilde{\phi}_{j\bm{k}}^{(q)}(\bm{r}+\bm{T},\sigma) = \widetilde{\phi}_{j\bm{k}}^{(q)}(\bm{r}\sigma),\label{Eq:2ndBloch}
\end{equation}
where $\bm{T}$ is the lattice translation vector. We will refer to the
dimensionless function $\widetilde{\phi}_{j\bm{k}}^{(q)}(\bm{r}\sigma)$ as a Bloch wave function.

If the potential is local in space, there is no correlation between unit
cells and the Hamiltonian can be written in a block diagonal form. It is
thus possible to regard the Bloch wave vector $\bm{k}$ as a sort of
quantum numbers associated with a translational symmetry in a broad sense.
That is, we can regard the Bloch wave vectors $\bm{k}$ and $\bar{\bm{k}}$
$(=-\bm{k})$ as the countable quantum numbers $\Omega$ and $\bar{\Omega}$,
respectively, which were introduced in Sec.~\ref{Sec:SLDAsymm}. We note
that the form of single-particle wave functions is the same for $\widetilde
{\phi}_{j\bar{\bm{k}}}^{(q)}(\bm{r}\sigma)$, meaning that the sign of the
exponent is different for $\widetilde{\phi}_{j\bar{\bm{k}}}$ because of
the relation, $\bar{\bm{k}}=-\bm{k}$. By substituting Eq.~\eqref{Eq:spwf_Bloch_k}
into Eqs.~\eqref{Eq:def_qpwf_u} and \eqref{Eq:def_qpwf_v}, we find that
the quasiparticle wave functions, $u_{\nu\bm{k}}(\bm{r}\sigma)$ and
$v_{\nu\bm{k}}(\bm{r}\sigma)$, can be written as follows:
\begin{eqnarray}
u_{\nu\bm{k}}^{(q)}(\bm{r}\sigma)
&=& \frac{1}{\sqrt{\mathcal{V}}}\widetilde{u}_{\nu\bm{k}}^{(q)}(\bm{r}\sigma)e^{\mathrm{i}\bm{k\cdot r}},\label{Eq:u_tilde}\\ 
v_{\nu\bm{k}}^{(q)}(\bm{r}\sigma)
&=& \frac{1}{\sqrt{\mathcal{V}}}\widetilde{v}_{\nu\bm{k}}^{(q)}(\bm{r}\sigma)e^{\mathrm{i}\bm{k\cdot r}},\label{Eq:v_tilde}
\end{eqnarray}
where $\widetilde{u}_{\nu\bm{k}}^{(q)}(\bm{r}\!+\!\bm{T},\sigma)=\widetilde{u}_{\nu\bm{k}}^{(q)}(\bm{r}\sigma)$
and $\widetilde{v}_{\nu\bm{k}}^{(q)}(\bm{r}\!+\!\bm{T},\sigma)=\widetilde{v}_{\nu\bm{k}}^{(q)}(\bm{r}\sigma)$
hold. Notice that the sign of the exponent is the same for both $u$ and $v$ components.
It is an important key to formulate the superfluid band theory as shown in the subsequent sections.

\begin{figure} [t]
\includegraphics[width=7.8cm]{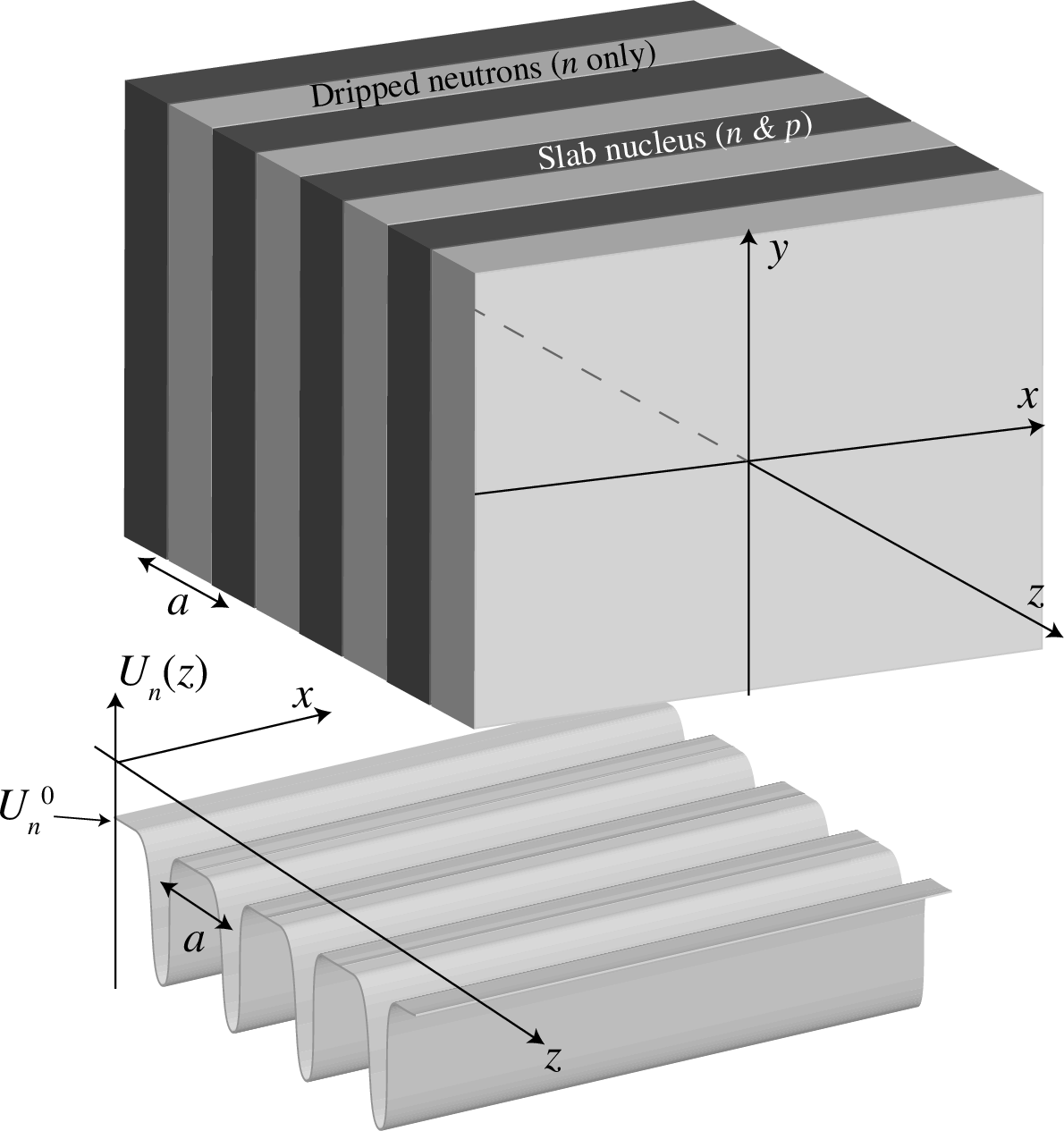}\vspace{2mm}
\caption{
Schematic picture showing the geometry of the systems under study. The nuclear slabs
extend parallel to $xy$ plane and are aligned with a period $a$ along $z$ direction,
as depicted in the upper part of the figure. In the lower part, the nuclear mean-field
potential is depicted in $xz$ plane. Note that when dripped neutrons exist the
maximum value of the mean field potential for neutrons ($U_{\rm n}^0$ in the picture)
is negative, even outside the slabs. The figure has been reprinted from Ref.~\cite{Kashiwaba(2019)}
with permission.
}
\label{Fig:slab}
\end{figure}

In the present paper, we consider 1-dimensional (1D) crystalline structure,
i.e.\ the slab (or ``lasagna'') phase of nuclear matter, where slabs are extending
parallel to $xy$ plane and are in a periodic sequence along $z$ direction;
See Fig.~\ref{Fig:slab}. In this case the lattice vector $\bm{T}$ reads
\begin{align}
\bm{T} = T_x\hat{\bm{e}}_x + T_y\hat{\bm{e}}_y + a n_z\hat{\bm{e}}_z,
\end{align}
where $T_x$ and $T_y$ are arbitrary real numbers, $n_z$ is an integer,
$a$ denotes the period of (or distance between) the neighboring slabs, and
$\hat{\bm{e}}_i$ is the unit vector along the $i$ ($=x$, $y$, or $z$) direction.
Since the single-particle wave functions along the $x$ and $y$ directions are
solely the plane waves, the Bloch's boundary condition, Eq.~(\ref{Eq:2ndBloch}),
is now reduced to
\begin{align}
\widetilde{\phi}^{(q)}_{j\bm{k}}(z+a,\sigma) = \widetilde{\phi}^{(q)}_{j\bm{k}}(z\sigma).
\end{align}
That is, the Bloch wave functions have 1D spatial dependence:
\begin{eqnarray}
u^{(q)}_{\nu\bm{k}}(\bm{r}\sigma) &=& \frac{1}{\sqrt{\mathcal{V}}}\widetilde{u}^{(q)}_{\nu\bm{k}}(z\sigma)e^{\mathrm{i}\bm{k}\cdot\bm{r}},\label{Eq:u_tilde(z)}\\[1mm]
v^{(q)}_{\nu\bm{k}}(\bm{r}\sigma) &=& \frac{1}{\sqrt{\mathcal{V}}}\widetilde{v}^{(q)}_{\nu\bm{k}}(z\sigma)e^{\mathrm{i}\bm{k}\cdot\bm{r}}.\label{Eq:v_tilde(z)}
\end{eqnarray}

The $z$ component of the Bloch wave vector can be reduced within
the first Brillouin zone, $-\pi/a\le k_z \le \pi/a$. In practical
calculations, we discretize the first Brillouin zone into $N_{k_z}$
points, i.e.\ $\Delta k_z = \frac{2\pi}{a}\frac{1}{N_{k_z}}$. This
implementation corresponds to a calculation with a length $\mathcal{L}
\equiv a N_{k_z}$ with the ordinary periodic boundary condition. Thus,
the normalization condition \eqref{Eq:normalization_qpwf} now reads
\begin{equation}
\mathcal{A}\,\sum_\sigma\int_0^{\mathcal{L}}\Bigl[ |u_{\nu\bm{k}}^{(q)}(\bm{r}\sigma)|^2+|v_{\nu\bm{k}}^{(q)}(\bm{r}\sigma)|^2 \Bigr]\dd z = 1,
\label{Eq:normalization_qpwf_L}
\end{equation}
where $\mathcal{A}$ stands for the normalization area such that
$\mathcal{V=LA}$. Because of the periodicity of the quasiparticle
wave functions, Eq.~\eqref{Eq:normalization_qpwf_L} is equivalent to
\begin{equation}
\sum_\sigma\int_0^a\Bigl[ |\widetilde{u}_{\nu\bm{k}}^{(q)}(z\sigma)|^2+|\widetilde{v}_{\nu\bm{k}}^{(q)}(z\sigma)|^2 \Bigr]\dd z = a.
\label{Eq:normalization_qpwf_a}
\end{equation}

\subsubsection{Energy density functionals}\label{Sec:EDF}

The main ingredient of DFT is the EDF. In this work, the EDF used
is almost the same as given in the previous work \cite{Sekizawa(2022)},
except the introduction of a pairing functional. For completeness,
here we briefly recall the equations, because some of them require
a caution specific to the superfluid systems.

The total energy per nucleon of the system is given by
\begin{equation}
\frac{E_\text{tot}}{A} = \frac{1}{N_\text{b}}
\int_0^a \bigl( \mathcal{E}_\text{nucl}(z) + \mathcal{E}_e(z) \bigr)\dd z,
\label{Eq:Etot/A}
\end{equation}
where $\mathcal{E}_\text{nucl}$ denotes a nuclear part of the energy density
and $\mathcal{E}_e$ is the electron's one. Here, $N_\text{b}=N_n+N_p$ [The
subscript `b' stands for `baryons' (which are nucleons, in the present
study)], where $N_q=\int_0^a n_q(z)\dd z$ is the total number of neutrons
($q=n$) or protons ($q=p$) per unit area within a single period $a$, with
$n_q(z)$ being the number density of neutrons ($q=n$) or protons ($q=p$).
The nuclear energy density is given as a sum of kinetic, nuclear (interaction),
and pairing energy densities,
\begin{equation}
\mathcal{E}_\text{nucl}(z) = \mathcal{E}_\text{kin}(z) + \mathcal{E}_\text{Sky}(z) + \mathcal{E}_\text{pair}(z),
\end{equation}
where
\begin{eqnarray}
\mathcal{E}_\text{kin}(z) &=& \sum_{q=n,p}\frac{\hbar^2}{2m_q}\tau_q(z),\\[1.5mm]
\mathcal{E}_\text{Sky}(z) &=& \sum_{t=0,1}\Big[ C^\rho_t[n_0]n^2_t(z) + C^{\Delta\rho}_t n_t(z)\partial_z^2 n_t(z) \nonumber\\[-1.5mm]
&&\hspace{8mm}+ C^\tau_t\bigl(n_t(z)\tau_t(z) - \bm{j}_t^2(z)\bigr) \Big],\label{Eq:Edns_Sky}\\[1.5mm]
\mathcal{E}_\text{pair}(z) &=& -\sum_{q = n,p} \Delta_q(z)\kappa_q^*(z),
\label{Eq:Edns_pair}
\end{eqnarray}
where $m_q$ is the nucleon mass\footnote{In this work,
$m_n=939.56542052$\,MeV/$c^2$ and $m_p=938.27208816$\,MeV/$c^2$
are used, as in Ref.~\cite{Sekizawa(2022)}.
} and $\partial_z$ represents a spatial derivative with respect to the
$z$ coordinate. In this work, we develop a formalism for a widely-used
Skyrme-type EDF for the nuclear part. The kinetic and momentum densities
in $\mathcal{E}_\text{Sky}$ \eqref{Eq:Edns_Sky} are formally defined,
respectively, by
\begin{eqnarray}
\tau_q(\bm{r}) &=& \bigl(\bm{\nabla\cdot\nabla}'\bigr)\,n_q(\bm{r},\bm{r}')\Bigr|_{\bm{r}=\bm{r}'},\label{Eq:def_tau}\\[2mm]
\bm{j}_q(\bm{r}) &=& \mathrm{Im}\bigl[(\bm{\nabla}-\bm{\nabla}')\,n_q(\bm{r},\bm{r}')\bigr]\Bigr|_{\bm{r}=\bm{r}'}, \label{Eq:def_j}
\end{eqnarray}
where $\bm{\nabla}$ and $\bm{\nabla}'$ act on the spatial coordinates
$\bm{r}$ and $\bm{r}'$, respectively. The time-odd momentum densities
vanish in static calculations, while they are, in general, finite in
a dynamic situation. The subscript $t$ in Eq.~\eqref{Eq:Edns_Sky}
specifies isoscalar ($t=0$) and isovector ($t=1$) densities, which are
defined, for the number density for instance, by $n_0(z)=n_n(z)+n_p(z)$
and $n_1(z)=n_n(z)-n_p(z)$, respectively (the subscript 0 is often omitted).
The detailed expressions of the coefficients by means of Skyrme force
parameters can be found in, e.g., Eq.~(A1) in Ref.~\cite{Lesinski(2007)}
(where symbols $A_t^\text{X}$ were used instead of $C_t^\text{X}$ here,
where X stands for $\rho$, $\tau$, and $\Delta\rho$). $C_t^\rho[n_0]$
depends on the local number density as $C_t^\rho[n_0]=C_t^\rho +
C_{t,\text{D}}^\rho n_0^\alpha(z)$ as in Ref.~\cite{Kortelainen(2010)}.

To evaluate the densities, Eqs.~\eqref{Eq:def_n}, \eqref{Eq:def_kappa},
\eqref{Eq:def_tau}, and \eqref{Eq:def_j}, we need to take summations over
all positive-energy quasiparticle states $\mu=\{\nu\,\bm{k}\}$. For the slab
phase under study, the summation over $k_x$ and $k_y$ can be replaced with
integrals, $\sum_{k_x,k_y}\rightarrow\int\mathcal{A}\, \dd k_x\dd k_y/(2\pi)^2
=\int\mathcal{A}\,k_\parallel\dd k_\parallel/(2\pi)$, where $k_\parallel
\equiv(k_x^2+k_y^2)^{1/2}$ is the magnitude of the Bloch wave vector parallel
to the slabs \cite{Sekizawa(2022)}. Then, the various densities can be
written as follows:
\begin{eqnarray}
n_q(z) &=& \sum_{\nu k_z\sigma}\int \frac{k_\parallel\dd k_\parallel}{2\pi\mathcal{L}} \big|v^{(q)}_{\nu\bm{k}}(z\sigma)\big|^2, \label{n_use}\\[1mm]
\tau_q(z) &=& \sum_{\nu k_z\sigma}\int \frac{k_\parallel\dd k_\parallel}{2\pi\mathcal{L}}
\Bigl[ k_\parallel^2 \big| v^{(q)}_{\nu\bm{k}}(z\sigma)\big|^2 \nonumber\\[-3mm]
&&\hspace{23mm}+ \big| (\partial_z +\mathrm{i}k_z)v^{(q)}_{\nu\bm{k}}(z\sigma)\big|^2 \Bigr], \label{tau_use}\\
\bm{j}_q(z) &=& -\sum_{\nu k_z\sigma}\int \frac{k_\parallel\dd k_\parallel}{2\pi\mathcal{L}} \nonumber\\
&&\times\Im\left[ v^{(q)*}_{\nu\bm{k}}(z\sigma)(\partial_z +\mathrm{i}k_z) v^{(q)}_{\nu\bm{k}}(z\sigma) \right]\hat{\bm{e}}_z, \label{j_use}\\[2mm]
\kappa_q(z) &=& \sum_{\nu k_z}\int \frac{k_\parallel\dd k_\parallel}{2\pi\mathcal{L}} v^{(q)*}_{\nu\bm{k}}(z\up)u^{(q)}_{\nu\bm{k}}(z\down). \label{K_use}
\end{eqnarray}
Note that a minus sign appears in the current density \eqref{j_use},
due to the definition \eqref{Eq:def_qpwf_v} that relates $v_{\nu\Omega}
(\bm{r}\sigma)$ and $\phi_{j\bar{\Omega}}^*(\bm{r}\sigma)$.

The pairing part of EDF, $\mathcal{E}_\text{pair}$ \eqref{Eq:Edns_pair},
contains the pairing field $\Delta_q(z)$ which is given by
\begin{equation}
\Delta_q(z) = -g_{q,\text{eff}}(z)\kappa_q(z),
\label{Eq:Delta_q}
\end{equation}
where $g_{q,\mathrm{eff}}(z)$ in Eq.~\eqref{Eq:Delta_q} is an effective pairing coupling constant
\cite{Bulgac(2002)2}, which is calculated as follows \cite{Jin(2021)}:
\begin{align}
\frac{1}{g_{q,\mathrm{eff}}(z)} = \frac{1}{g_0} - \frac{m^{\oplus}_q(z)}{4\pi^2\hbar^2}\frac{\pi}{a}K,
\end{align}
where $g_0$ is the bare coupling constant, $m_q^\oplus(z)$ is a ``microscopic''
effective mass, which will be defined in Eq.~\eqref{Eq:def_micro_m*}, and $K$
is a numerical constant that corresponds to the integral,
\begin{align}
K &= \frac{12}{\pi}\int_0^{\!\frac{4}{\pi}} \ln(1+1/\cos^2\theta) \dd\theta\nonumber\\
  & = 2.442749607806335\dots.
\end{align}
We set the bare coupling constant to $g_0=200$\;MeV\,fm$^3$ in the present work.
This value has been used for finite nuclei in the literature \cite{Bulgac(2002),Yu(2003),Bulgac(2018)}.
We find that this coupling constant yields a reasonable value of the pairing field for neutron-star
matter on the order of 1\,MeV [cf. Fig.~\ref{betanb05}(b)]. In the present work, we thus employ
this coupling constant for the sake of simplicity. It is to mention that there exist several
refined treatments of the coupling constant that correctly reproduce density dependence of
the neutron pairing gap in neutron star matter (see, e.g., Refs.~\cite{Wlazlowski(2016),Okihashi(2021)})

The Coulomb part of EDF reads
\begin{eqnarray}
\mathcal{E}_{\mathrm{Coul}}^{(p)}(z) &=& \frac{1}{2}n_p(z)V_{\mathrm{Coul}}(z)
- \frac{3e^2}{4}\biggl(\frac{3}{\pi}\biggr)^{\!\!1/3}\!\!\!n_p^{4/3}(z),
\end{eqnarray}
where $V_{\mathrm{Coul}}(z)$ denotes the Coulomb potential for protons and
$e$ is the elementary charge. The Slater approximation has been adopted
for the Coulomb exchange functional. For the Coulomb potential, we solve
the Poisson equation,
\begin{align}
    \frac{\dd^2}{\dd z^2}V_{\mathrm{Coul}}(z) = -\frac{e^2}{\varepsilon_0}n_{\mathrm{ch}}(\bm{r}),
\end{align}
where $\varepsilon_0$ is the vacuum permittivity. Here, $n_{\mathrm{ch}}(z)
\equiv n_p(z) - n_e$ denotes the charge density, neglecting the charge
form factor of protons. Electrons are assumed to be uniformly distributed
with the density $n_e = \bar{n}_p$, where $\bar{n}_q=\frac{1}{a}\int_0^a n_q(z)\dd z$
is the average nucleon number density. The Coulomb potential is subjected to the
charge neutrality condition, $\int_0^a V_\text{Coul}(z)\dd z=0$.

For the electron's EDF, $\mathcal{E}_e(z)$ in Eq.~\eqref{Eq:Etot/A}, we use
formulas for a relativistic electron gas. For explicit expressions, We refer
readers to Refs.~\cite{Kashiwaba(2019),Sekizawa(2022)}.

\subsubsection{Skyrme (TD)SLDA equations for the slab phase}\label{Sec:SkyrmeHFB}

From an appropriate functional derivative, one can derive the corresponding
single-particle Hamiltonian which enters the (TD)SLDA equation. Because our
working EDFs are local in space, the resulting equation becomes also a local one.
In the same way as the normal (without pairing correlations) self-consistent
band theory \cite{Sekizawa(2022)}, the point is that an operation of a spatial
derivative on a quasiparticle wave function \eqref{Eq:u_tilde(z)} generates
an additional $\bm{k}$-dependent term as follows:
\begin{equation}
\bm{\nabla}u_{\nu\bm{k}}^{(q)}(\bm{r}\sigma) = \frac{1}{\sqrt{\mathcal{V}}}
e^{\mathrm{i}\bm{k\cdot r}} \bigl(\partial_z\hat{\bm{e}}_z+\mathrm{i}\bm{k}\bigr)
\widetilde{u}_{\nu\bm{k}}^{(q)}(z\sigma),
\end{equation}
where the same is true also for the $v$ component \eqref{Eq:v_tilde(z)}.
Substituting Eqs.~\eqref{Eq:u_tilde(z)} and \eqref{Eq:v_tilde(z)} into
a localized version of Eq.~\eqref{Eq:HFB_coordinate-space_reduced}, and
factoring out the common function $e^{\mathrm{i}\bm{k\cdot r}}$ after
operations of the spatial derivatives, we obtain
\begin{align}
\mqty(\hat{h}^{(q)}(z)+\hat{h}^{(q)}_{\bm{k}}(z)-\lambda_q & \Delta_q(z)\\
\Delta_q^*(z)              & -\hat{h}^{(q)*}(z)-\hat{h}^{(q)*}_{-\bm{k}}(z)+\lambda_q) \nonumber\\[1mm]
\times
\mqty(\widetilde{u}^{(q)}_{\nu\bm{k}}(z\up) \\
\widetilde{v}^{(q)}_{\nu\bm{k}}(z\down))
= E_{\nu\bm{k}}
\mqty(\widetilde{u}^{(q)}_{\nu\bm{k}}(z\up) \\
\widetilde{v}^{(q)}_{\nu\bm{k}}(z\down)).
\end{align}
It should be noted that, while we deal with a three-dimensional system,
the equations to be solved are essentially one-dimensional ones, significantly
reducing the computational cost.

The single-particle Hamiltonian, $\hat{h}^{(q)}(z)$, is given by
\begin{align}
    \hat{h}^{(q)}(z) = &-\div M^{(q)}(z)\grad + U^{(q)}(z)\nonumber\\
    &+ \frac{1}{2\mathrm{i}}\left[ \div \bm{I}^{(q)}(z) + \bm{I}^{(q)}(z)\bm{\cdot}\grad \right].\label{Hamil1}
\end{align}
Note that the differential operators in Eq.~\eqref{Hamil1} act on all spatial
functions located right side of them. Here, we have introduced a time-even
mean-field $M^{(q)}(z)$ defined as
\begin{align}
M^{(q)}(z)\equiv \frac{\hbar^2}{2m^\oplus_q(z)} = \frac{\hbar^2}{2m_q} + \sum_{q^\prime = n, p}C^{\tau(q)}_{q^\prime}n_{q^\prime}(z),
\label{Eq:def_micro_m*}
\end{align}
where $m^{\oplus}_q(z)$ is the ``microscopic'' effective mass which should be
distinguished from a ``macroscopic'' one discussed in Sec.~\ref{Sec:anti-entrainment}.
$U^{(q)}(z)$ and $\bm{I}^{(q)}(z)$ are time-even and time-odd mean field potentials,
respectively, defined as
\begin{align}
    U^{(q)}(z) =& \sum_{q^\prime = n, p}\Big[ 2C^{\rho(q)}_{q^\prime}n_{q^\prime}(z) + 2C^{\nabla\rho(q)}_{q^\prime}\partial^2_z n_{q^\prime}(z) \nonumber\\
    & + C^{\tau(q)}_{q^\prime}\tau_{q^\prime}(z) + 2n_0^\alpha(z)C^{\rho(q)}_{q^\prime D}n_{q^\prime}(z) \Big]\nonumber\\
    &+ \alpha n_0^{\alpha-1}(z)\sum_{t = 0,1}C^\rho_{tD}n^2_t(z)\nonumber\\
    &+ U_{\mathrm{Coul}}(z)\delta_{qp}\nonumber\\[1mm]
    &+ \sum_{q^\prime = n,p} \frac{\partial g_{q'\!,\text{eff}}}{\partial n_q}|\kappa_{q^\prime}(z)|^2\label{time_even_mean_potential},\\[1mm]
    \bm{I}^{(q)}(z) &= -2 \sum_{q^\prime = n, p}C^{\tau(q)}_{q^\prime}\bm{j}_{q^\prime}(z),
\end{align}
where 
\begin{align}
    U_{\mathrm{Coul}}(z) = V_{\mathrm{Coul}}(z) - e^2 \left( \frac{3}{\pi} \right)^{\!\!1/3}\!\!\!n_p^{1/3}(z).
\end{align}
Note that there is an additional contribution to $U^{(q)}(z)$ arising from the
density derivative of the effective pairing coupling constant, which is given by
\begin{equation}
\pdv{g_{q'\!,\text{eff}}}{n_q}
= \bigl[g_{q'\!,\mathrm{eff}}(z)\bigr]^2\frac{K}{8\pi a}
\Biggl(\!\frac{\hbar^2}{2m^{\oplus}_{q'}(z)}\!\Biggr)^{\!\!\!-2}\!\!C^{\tau(q^\prime)}_{q}.
\end{equation}
Following the previous work \cite{Sekizawa(2022)}, we have defined
a shorthand notation,
\begin{eqnarray}
C_n^{\text{X}(q)} &\equiv& C_0^\text{X} + \eta_q C_1^\text{X},\\[1.5mm]
C_p^{\text{X}(q)} &\equiv& C_0^\text{X} - \eta_q C_1^\text{X},
\end{eqnarray}
where X stands for the superscript of the coefficients, i.e., $\rho$,
$\tau$, or $\Delta\rho$, and $\eta_q=+1$ ($-1$) for neutrons (protons).
We note that the time-odd potential vanishes in a static situation. The
single-particle Hamiltonian which depends on the Bloch wave vector,
$\hat{h}^{(q)}_{\bm{k}}(z)$, can be represented as follows \cite{Sekizawa(2022)}:
\begin{align}
\hat{h}^{(q)}_{\bm{k}}(z) = \frac{\hbar^2\bm{k}^2}{2m^{\oplus}_q(z)} + \hbar \bm{k}\bm{\cdot}\hat{\bm{v}}^{(q)}(z),\label{H_k}
\end{align}
where $\hat{\bm{v}}^{(q)}(z)$ is the so-called velocity operator,
\begin{align}
    \hat{\bm{v}}^{(q)}(z) &= \frac{1}{\mathrm{i}\hbar}\left[ \bm{r}, \hat{h}^{(q)}(z) \right]\nonumber\\
    &= -\mathrm{i}\hbar \left( \grad\frac{1}{2m^{\oplus}_q(z)} + \frac{1}{2m^{\oplus}_q(z)}\grad  \right) + \frac{1}{\hbar} \bm{I}_q(z).
\end{align}

As will be described in Sec.~\ref{Sec:anti-entrainment}, we apply the
real-time method, proposed in Ref.~\cite{Sekizawa(2022)}, where we extract
the collective masses of a slab and of protons from a dynamic response
of the system to an external force. The external force can be introduced by
means of a time-dependent, uniform vector potential $A_z(t)$ that couples only
with protons which are localized inside slabs. Such a vector potential is
equivalent to a uniform electric field, $E_z(t)=-(1/c)\dd A_z(t)/\dd t$.
Time evolution of the system is described by the TDSLDA equation in the
velocity gauge \cite{Sekizawa(2022)}:
\begin{align}
\mqty(\hat{h}^{(q)}(z,t)+\hat{h}^{(q)}_{\bm{k}(t)}(z,t) & \Delta_q(z,t)\\
\Delta_q^*(z,t)              & -\hat{h}^{(q)*}(z,t)-\hat{h}^{(q)*}_{-\bm{k}(t)}(z,t))\nonumber\\
\times\mqty(\widetilde{u}^{\prime(q)}_{\nu\bm{k}}(z\up,t) \\
\widetilde{v}^{\prime(q)}_{\nu\bm{k}}(z\down,t))
\;=\;
\mathrm{i}\hbar\frac{\partial}{\partial t}
\mqty(\widetilde{u}^{\prime(q)}_{\nu\bm{k}}(z\up,t) \\
\widetilde{v}^{\prime(q)}_{\nu\bm{k}}(z\down,t)),
\label{Eq:TDSLDA-v}
\end{align}
where the prime on the quasiparticle wave functions indicates that they are
represented in the velocity gauge,
\begin{eqnarray}
\widetilde{u}_{\nu\bm{k}}^{\prime(q)}(z,t)
\equiv \exp\biggl[ -\frac{\mathrm{i}e}{\hbar c}A_z(t)z \biggr] \widetilde{u}_{\nu\bm{k}}^{(q)}(z,t),
\end{eqnarray}
and a similar formula (with the opposite sign in the exponent) holds for the $v$
components \eqref{Eq:v_tilde(z)}. Notice that the Bloch wave vector in the $\bm{k}$-dependent
single-particle Hamiltonian \eqref{H_k} in Eq.~\eqref{Eq:TDSLDA-v} is shifted as a function of
time, according to the following relation:
\begin{equation}
\bm{k}(t) = \bm{k} + \frac{e}{\hbar c}A_z(t)\hat{\bm{e}}_z.
\end{equation}
All densities can be expressed in terms of the quasiparticle wave functions
in the velocity gauge, replacing $\widetilde{u}_{\nu\bm{k}}^{(q)}\rightarrow
\widetilde{u}_{\nu\bm{k}}^{\prime(q)}$, $\widetilde{v}_{\nu\bm{k}}^{(q)}
\rightarrow\widetilde{v}_{\nu\bm{k}}^{\prime(q)}$, and $k_z\rightarrow k_z(t)$.
More detailed explanations on the expressions in the velocity gauge can be
found in Ref.~\cite{Sekizawa(2022)}.

\section{Computational Details}\label{Sec:ComptDetails}

We have newly developed a parallel computational code from scratch that reproduces
all the results presented in Ref.~\cite{Sekizawa(2022)} and extended it to include
superfluidity. All the calculations were carried out with Skyrme SLy4 EDF
\cite{Chabanat(1998)} as in Ref.~\cite{Sekizawa(2022)}. We consider a situation
where nuclear slabs extend along $xy$ directions, forming a perfect crystalline
structure along $z$ axis with a period $a$. We discretize the $z$ coordinate into
a uniform grid with spacing $\Delta z$ to represent qusiparticle wave functions.
The mesh spacing is set to $\Delta z=0.5$\,fm. The spectral method with fast
Fourier transformations (FFTs) are used to evaluate the first and second spacial derivatives.
The Poisson equation for the Coulomb potential is solved also with the FFT algorithm.
We discretize the first Brillouin zone $-\pi/a \leq k_z \leq \pi/a$ into $N_{k_z}$
points. We use $N_{k_z}=80$ as in Ref.~\cite{Sekizawa(2022)}. For the Bloch wave
vector parallel to the slabs, $k_\parallel$, we introduce a cutoff $k^{\mathrm{max}}
_{\parallel}$ and discretize it with a $\Delta k_\parallel$ step. For the calculations
presented in this paper, $k^{\mathrm{max}}_\parallel = 1.5$\,fm$^{-1}$ and
$\Delta k_\parallel = 0.01$\,fm$^{-1}$ are adopted. We have confirmed that
these computational settings provide satisfactory convergent results for the
systems under study.

For time evolution, we use the eighth-order Taylor expansion method with a single
predictor-corrector step with $\Delta t = 0.1$\,fm/$c$. With this setting, we have
confirmed that the total number of nucleons and the total energy per nucleon are
conserved with $10^{-6}$- and $10^{-10}$-MeV accuracy, respectively, within the
simulation time of 4,000\,fm/$c$. For the extraction of the collective mass of
a slab, we dynamically introduce an external potential for protons, $E_z(t) =
S(\eta,w,t)E_z$, as a function of time $t$, where 
\begin{align}
    S(\eta,w,t) = \frac{1}{2} + \frac{1}{2}\tanh\left[\eta\tan(\frac{\pi t}{w} - \frac{\pi}{2})\right]
\end{align}
is a switching function which varies smoothly from $0$ to $1$ within an interval $t=[0,w]$.
In this way we can avoid unnecessary excitations of the system \cite{Sekizawa(2022)}.
In the present paper, $\eta=1$ and $w=\text{2,000}$\,fm/$c$ were used and calculations
were continued up to $t=\text{4,000}$\,fm/$c$. We set the strength of the external
potential as $eE_z=10^{-3}$\,MeV/fm, as in Ref.~\cite{Sekizawa(2022)}.

The slab period $a$ under study is about $a\approx30$\,fm and, thus, the number
of grid points along the $z$ coordinate is $N_z\approx60$. In such a case, the
total number of quasiparticle wave functions that need to be solved in the
current computational setup is estimated to be
\begin{eqnarray}
&N_z\times N_{k_z} \times N_{k_\parallel} \times \text{2 (isospin)} \times \text{2 ($u$ and $v$)}& \nonumber\\[2mm]
&= 60\times 80\times 150\times 4 = \text{2,880,000}.& \nonumber
\end{eqnarray}
That is, although 1D equations look easily tractable, one has to deal
with millions of complex, non-linear, partial differential equations
for the quasiparticle wave functions. Note that if the single-particle
Hamiltonian involves a term that mixes spins the number increases
by a factor of two, because of an explicit treatment for the spin
degree of freedom. Currently, our code is parallelized with respect
to the Bloch wave number $k_z$ using the message passing interface (MPI).

\section{Results and Discussion}\label{Sec:Results}

\subsection{On the convergence of self-consistent calculations}

To solve the static SLDA equation, we performed iterative diagonalizations of
the Hamiltonian matrix. At every step $m$, the chemical potentials for neutrons
and protons $\lambda_q$ were updated to obtain correct neutron and proton numbers
satisfying the $\beta$-equilibrium condition. To this end, we adjust the chemical
potential of protons as $\lambda^{(m+1)}_p = \lambda_p^{(m)} - \alpha_\lambda(N_\text{b}^{(m+1)}
- N^{(0)}_\text{b})$ with $\alpha_\lambda = 50$, where $N_\text{b}^{(0)}$ denotes
the requested nucleon number, i.e., $N_\text{b}^{(0)}=an_\text{b}$. While, the chemical
potential of neutrons are determined simply through the $\beta$-equilibrium condition,
$\mu_n=\mu_p+\mu_e$. From the calculations, we realized that a use of the modified
Broyden's method (see, e.g., Ref.~\cite{Baran(2008)} and references therein)
is crucial to get a convergent result with a reasonable computational time. The
modified Broyden mixing was applied to the mean-field, pairing, and chemical
potentials so that the unitarity of the Bogoliubov transformation matrix is always
preserved \cite{Baran(2008)}. The Broyden vector thus reads $\boldsymbol{V}=
\{M^{(n)}(z),M^{(p)}(z),U^{(n)}(z),U^{(p)}(z), \mathrm{Re}[\Delta_n(z)],$
$\mathrm{Im}[\Delta_n(z)],\mathrm{Re}[\Delta_p(z)],\mathrm{Im}[\Delta_p(z)],
\mu_n,\mu_p\}$, i.e., with $8N_z+2$ dimensions. The modified Broyden's method
also contains a parameter $\alpha$ that controls a rate of mixing, which we set $\alpha=0.1$.

\begin{figure}[t]
    \centering
    \includegraphics[width=\columnwidth]{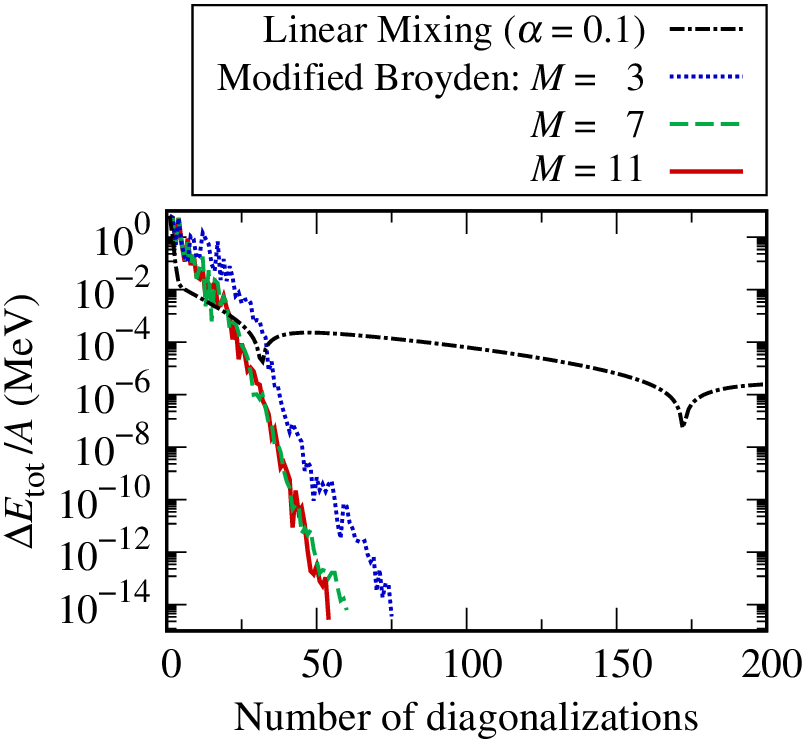}
    \caption{
    The change of the total energy per nucleon between two successive iterations,
    $\Delta E_\text{tot}/A=|E_\text{tot}^{(m)}-E_\text{tot}^{(m-1)}|/A$, is plotted as
    a function of the number of diagonalizations $m$ for the system with $n_\text{b}=0.05$\,fm$^{-3}$
    under the $\beta$-equilibrium condition. The results obtained with the modified
    Broyden's method with $M=11$, 7, and 3 are shown by solid, dashed, and dotted lines,
    respectively. The result obtained with the linear mixing method with $\alpha=0.1$
    is also shown by a dot-dashed line for comparison.
    }
    \label{Broyden}
\end{figure}

As a typical example, we show in Fig.~\ref{Broyden} change in total energy
per nucleon, $\Delta E_\text{tot}/A\equiv|E_\mathrm{tot}^{(m)}-E_\mathrm{tot}^{(m-1)}|/A$,
as a function of the number of diagonalizations for $n_\mathrm{b}=0.05$\,fm$^{-3}$
under the $\beta$-equilibrium condition. In the modified Broyden's method,
information of $M$ previous steps is used for updating the Broyden vector.
For comparison, results with $M=11$, 7, and 3 are shown by solid, dashed,
and dotted lines, respectively. A result obtained with a simple linear mixing
is also presented by a dash-dotted line. In the linear mixing method, potentials were updated simply
according to $X^{(m+1)}=(1-\alpha)X^{(m)} + \alpha X'^{(m+1)}$ with $\alpha=0.1$,
where $X^{\prime(m+1)}$ stands here for a tentative potential obtained after the
($m$\,$+$\,$1$)th diagonalization of the Hamiltonian matrix. From the figure, one finds that
the linear mixing may be good for the first 10--20 iterations, but its convergence
is very slow, which cannot reach $\Delta E<10^{-6}$\,MeV even after 200 diagonalizations.
One may naively expect that the convergence would be improved by increasing the
rate of mixing, however, we found that the calculation becomes unstable already
for $\alpha=0.15$. In contrast, we find a good convergent behavior with the modified
Broyden's method, where the energy change decreases exponentially down to $10^{-14}$\,MeV.
In the present paper, we use $M=7$ and set a convergence criterion as
$\Delta E_\text{tot}/A<10^{-10}$\,MeV.

\subsection{On the optimal slab period}\label{Sec:a_opt}

In self-consistent band theory calculations, the slab period $a$
is linked to the size of the computational region, $N_z\Delta z$.
Therefore, to figure out the optimal slab period that minimizes the
total energy of the system, we need to repeat static calculations
by changing the number of latiice points, $N_z$.

\begin{figure}[t]
    \centering
    \includegraphics[width=0.92\columnwidth]{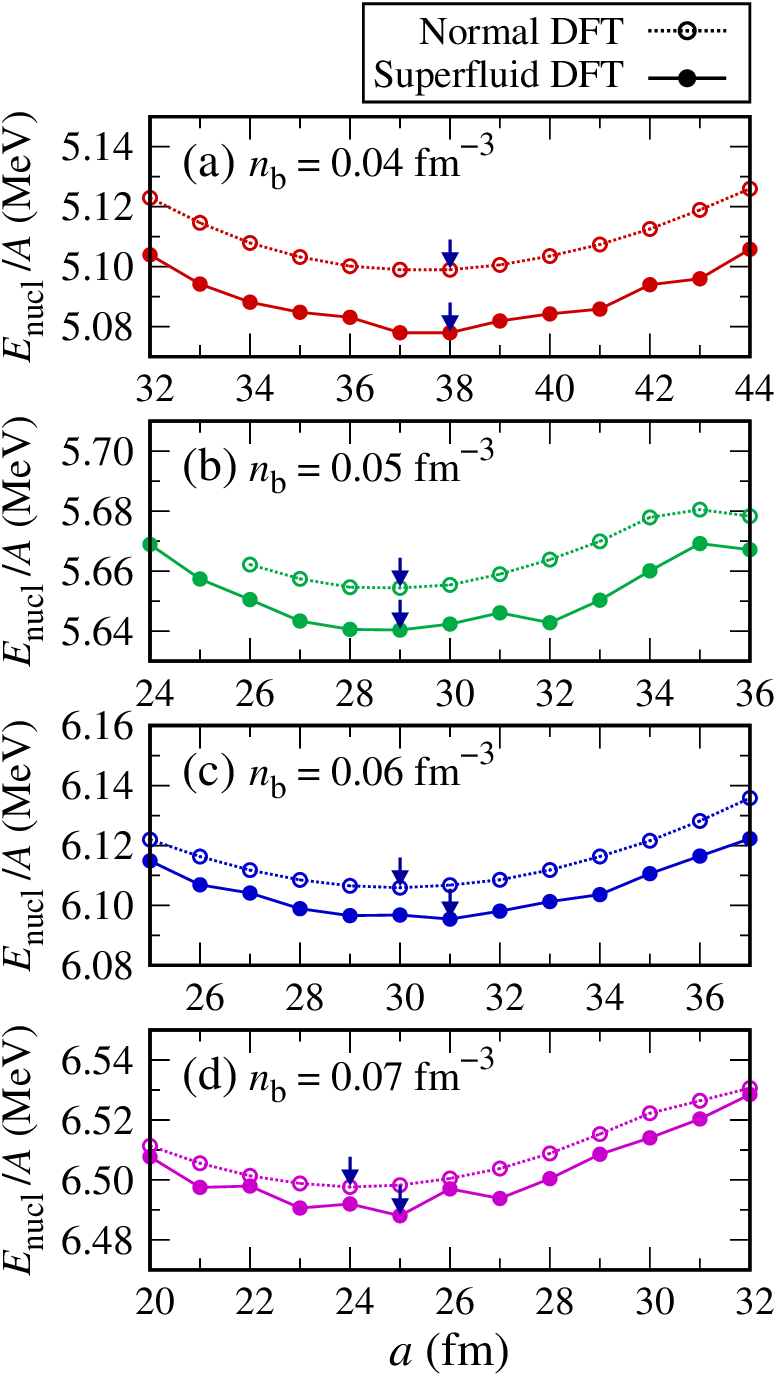}
    \caption{
    Nuclear energy per nucleon, $E_\text{nucl}/A$, is shown as a function of the
    slab period $a$. Results for $n_\text{b}=0.04$, 0.05, 0.06, and 0.07\,fm$^{-3}$
    are exhibited in panels (a), (b), (c), and (d), respectively. In each panel,
    results of superfluid band calculations are shown by filled circles connected
    with solid lines, while those without superfluidity are represented by
    open circles connected with dotted lines. The minimum energy locations
    are indicated by arrows.
    }
    \label{AvsE_nb05}
\end{figure}

To show how the energy depends on the slab period $a$, we show in Fig.~\ref{AvsE_nb05}
the nuclear energy per nucleon, i.e.\ $E_\text{nucl}\equiv\frac{1}{N_\text{b}}
\int_0^a \mathcal{E}_\text{nucl}(z)\dd z$, as a function of the slab period $a$ for
$n_\text{b}=0.04$, 0.05, 0.06, and 0.07\,fm$^{-3}$ under the $\beta$-equilibrium
condition. The results of superfluid band theory calculations
are shown by solid circles connected with solid lines, while those of
normal band theory (without pairing correlations) are shown by open
circles connected with dotted lines. From the figure, we find that the
total energy is always lower with superfluidity, gaining energy through
the pairing correlations, as expected for the inner crust of neutron stars.
In addition, a parabolic behavior is visible, which is associated with
balance between Coulomb and nuclear interactions. While a too small
slab period is energetically unfavorable due to the Coulomb repulsion,
a too large value results in loss of nuclear attraction. In the figure,
the optimal slab period that minimizes the total energy is indicated
by an arrow. It is visible from Figs.~\ref{AvsE_nb05}(c) and \ref{AvsE_nb05}(d)
that inclusion of superfluidity could slightly affect the period $a$.

It should be noted that the energy curve shows somewhat different behavior
when one includes electrons' contribution. Namely, the total energy per nucleon,
$E_\text{tot}/A$, exhibits a gentle dependence on the slab period $a$ with smaller
curvature, and the resulting optimal slab period tends to be larger than those
shown in Fig.~\ref{AvsE_nb05} (see Appendix~\ref{App:Etot/A} for details).
It is, of course, desirable to optimize the slab period $a$ by minimizing
the total energy of the system. In the present work, however, we analyze the
systems that minimize the nuclear energy, following Ref.~\cite{Kashiwaba(2019)},
because the primary purpose of this study is to quantify the effects of band
structure and superfluidity for systems with different baryon number densities.
In the future works, where we extend the theoretical framework to deal with 2D
and 3D crystalline structures and make a quantitative prediction on the neutron
effective mass, configurations that minimize the total energy should be investigated.

\begin{table}[t]
    \centering
    \caption{
    A summary of self-consistent superfluid band theory calculations for
    a range of baryon number densities $n_\text{b}=0.04$--$0.07$\,fm$^{-3}$ under
    the $\beta$-equilibrium condition. From the left to right columns,
    the baryon number density, $n_\text{b}$, in fm$^{-3}$, proton fraction,
    $Y_p$, optimal slab period, $a$, in fm, background neutron number density,
    $n_n^\text{bg}$, in fm$^{-3}$, and average absolute values of the
    pairing field, $\overline{\Delta}_{q}$ ($q=n$ for neutrons and $q=p$
    for protons), in MeV, are presented.
    }\vspace{3mm}
    \begin{tabular*}{\columnwidth}{@{\extracolsep{\fill}}cccccc}
        \hline\hline$n_\text{b}$ & $Y_p$ & $a$ & $n_n^\text{bg}$ & $\overline{\Delta}_{n}$ & $\overline{\Delta}_{p}$ \\
        \hline$0.04$ & $3.31\times 10^{-2}$ & $38$ & $3.23\times 10^{-2}$ & $1.20$ & $9.34\times 10^{-2}$ \\
        $0.05$ & $3.30\times 10^{-2}$ & $29$ & $4.08\times 10^{-2}$ & $1.34$ & $6.88\times 10^{-2}$ \\
        $0.06$ & $3.37\times 10^{-2}$ & $31$ & $4.96\times 10^{-2}$ & $1.49$ & $8.33\times 10^{-2}$ \\
        $0.07$ & $3.50\times 10^{-2}$ & $25$ & $5.80\times 10^{-2}$ & $1.55$ & $2.25\times 10^{-2}$ \\ 
        \hline\hline
    \end{tabular*}
    \label{BetaResultSummary}
\end{table}

\subsection{Static properties and band structure}

The results of fully self-consistent superfluid band theory calculations are
summarized in Table.~\ref{BetaResultSummary}. In the table, we show
the proton fraction, $Y_p$, which is determined by the $\beta$-equilibrium
condition, the optimal slab period, $a$, the background neutron number
density, $n_n^{\text{bg}}$, and the average absolute value of the pairing
field, $\overline{\Delta}_q$. Clearly, the system is quite neutron-rich,
with proton fractions around $0.033$--$0.035$. As baryon number density
increases, the optimal slab period $a$ decreases, as expected. In our calculations,
a nuclear slab is located at the center of the computational region, $z=a/2$,
and the background neutron number density is defined as $n_n^\text{bg}=n_n(0)
=n_n(a)$. The background neutron number density $n_n^\text{bg}$ increases
with density, indicating that more neutrons are dripped outside of the slab for
higher densities. The average absolute values of the pairing field is defined by
\begin{equation}
\overline{\Delta}_q = \frac{1}{N_q}\int_0^a |\Delta_q(z)|n_q(z)\dd z.
\label{Eq:average_Delta_q}
\end{equation}
In the present computational settings, both neutrons and protons are
found to be in the superfluid and superconducting phases, respectively,
although the average proton pairing gap is very small, as compared to
the neutron's one.

\begin{figure}[t]
    \centering
    \includegraphics[width=0.95\columnwidth]{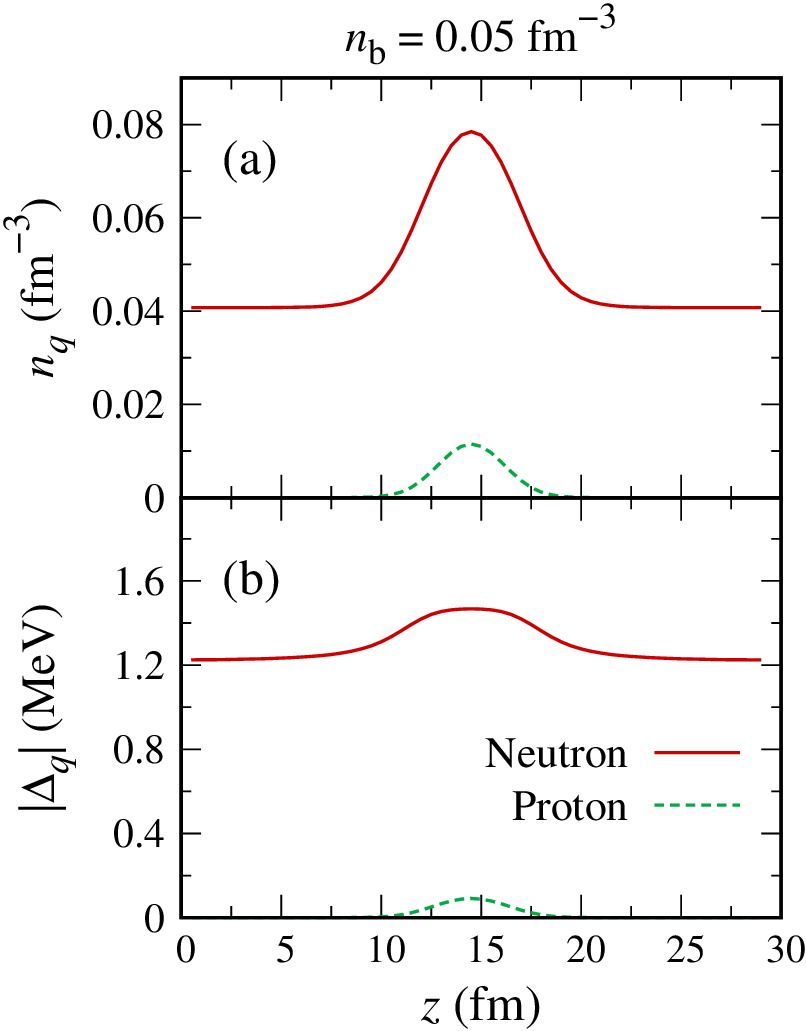}
    \caption{
    Nucleon number densities, $n_q(z)$, and the absolute value of the pairing field,
    $|\Delta_q(z)|$, for the system with $n_\text{b}=0.05$\,fm$^{-3}$ under the
    $\beta$-equilibrium condition are shown in panels (a) and (b), respectively,
    as a function of $z$ coordinate. Solid lines show those of neutrons, while
    dashed lines show those of protons.
    }
    \label{betanb05}
\end{figure}

In Fig.~\ref{betanb05}, we show a typical density distribution (a) and
the absolute value of the pairing field (b) for the system with $n_\text{b}=
0.05$\,fm$^{-3}$ under the $\beta$-equilibrium condition. Because of the small
proton fraction $Y_p=0.033$, there are substantial portion of dripped neutrons
outside of the slab, whose density distribution looks quite diffusive (or
``melting,'' close to the uniform nuclear matter). This characteristic is
more pronounced in the case of distribution of the pairing field. The
absolute value of the pairing field is much larger for neutrons, but
protons are also superconducting although $\overline{\Delta}_p$ is small.
Since substantial amount of neutrons are dripped outside of the slab,
we can expect the formation of band structure. Now, a question arises:
how the superfluidity affects the band structure?

\begin{figure}[t]
    \centering
    \includegraphics[width=0.95\columnwidth]{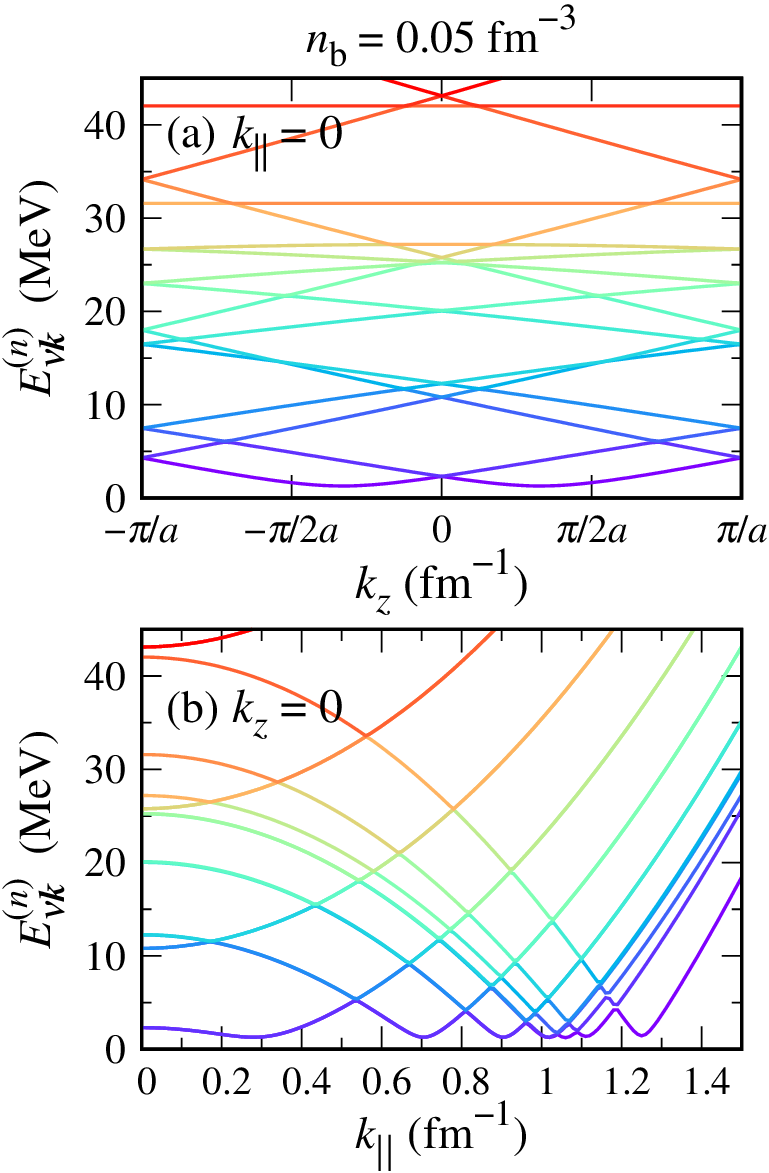}
    \caption{
    Quasiparticle energies of neutrons, $E_{\nu\bm{k}}^{(n)}$, are shown as functions
    of $k_z$ for $k_\parallel=0$ (a) and as functions of $k_\parallel$ for $k_z=0$ (b)
    for the system with $n_\text{b}=0.05$\,fm$^{-3}$ under the $\beta$-equilibrium
    condition. Line color changes gradually according to the ascending order of
    $E_{\nu\bm{k}}^{(n)}$, just to guide the eye.
    }
    \label{Energystates_nb05}
\end{figure}

Figure~\ref{Energystates_nb05} shows the quasiparticle energies $E_{\nu\bm{k}}^{(q)}$
for the system with $n_\text{b}=0.05$\,fm$^{-3}$ under the $\beta$-equilibrium
condition, as an illustrative example. In Fig.~\ref{Energystates_nb05}(a),
the results are plotted as a function of the Bloch wave number $k_z$ within
the first Brillouin zone, $-\pi/a\le k_z\le \pi/a$, with $k_\parallel=0$,
while quasiparticle energies for $k_z>0$ are plotted as a function of $k_\parallel$
in Fig.~\ref{Energystates_nb05}(b). Line color changes gradually according to
the ascending order of $E_{\nu\bm{k}}^{(n)}$, just to guide the eye. One
should keep in mind that quasiparticle energies are related to single-particle
ones in the canonical basis as
\begin{equation}
E_\mu = \pm\sqrt{(\varepsilon_\mu-\lambda)^2 + \Delta^2}.
\end{equation}
From Fig.~\ref{Energystates_nb05}(a), deep-hole states which are within
the potential well, $\varepsilon_\mu<U_n^0$, can be seen as horizontal
lines above 30\,MeV\footnote{$U_n^0$ denotes the maximum value of the mean-field
potential, which takes negative values even outside of the slabs when dripped neutrons
exist (cf. Fig.~\ref{Fig:slab}).}. In the figure, many other lines are visible,
showing $k_z$ dependence, which are associated with dripped neutrons that
extend spatially outside the slabs. In Fig.~\ref{Energystates_nb05}(b),
on the other hand, we find there are parabolic curves, some of which
are convex upward and the rest is opposite. Basically, $k_\parallel$
dependence originates from $\hbar^2 k_\parallel^2/(2m_q^\oplus)$ in Eq.~\eqref{H_k}
and, thus, those curves which are convex upward are contributions solely
from states below the chemical potential. Clearly, quasiparticle energies
exhibit complex dependence on the Bloch wave vector $\bm{k}$, which we call
the band structure. We will discuss the corresponding single-particle
energies in Sec.~\ref{Sec:qpresonances} (cf.~Fig.~\ref{res_nb05}).

\subsection{Anti-entrainment effects}\label{Sec:anti-entrainment}

\subsubsection{Real-time method}

In this section, we present results of fully self-consistent time-dependent
simulations to quantify the entrainment effects in the slab phase of neutron-star
matter in the presence of superfluidity. We employ a real-time method that enables
us to extract the collective masses of a slab and of protons from a dynamic response
of the system to an external force, the method proposed in Ref.~\cite{Sekizawa(2022)}.
Here we succinctly digest the essence of the approach, referring readers to
Ref.~\cite{Sekizawa(2022)} for details.

In the inner crust of neutron stars, where dripped superfluid neutrons
permeate a crystalline structure of nuclear bundles, it is not at all
obvious how to distinguish bound and unbound (free) neutrons. One may
naively subtract background uniform neutron density to define density
of a cluster as a ``bump'' in the whole density distribution, or compare
depth of a mean-field potential and single-particle energies to count
the number of neutrons within the potential well \cite{Papakonstantinou(2013)}.
In real situations, however, due to the self-organizing character of
neutron-star matter, i.e.\ there is no clear-cut separation between
clusters and the others, those are served only as a naive estimation.
In addition, part of dripped neutrons are expected to be immobilized
or trapped by the periodic structure through the Bragg scattering.
This is the so-called entrainment effect and the amount of the
``effectively-bound'' neutrons is the matter of debate. We note,
moreover, that superfluidity may also affect the collective mass of
nuclear clusters \cite{Martin(2016)}. It is thus a quite complicated
problem and only fully self-consistent superfluid band calculations
developed here can provide conclusive results.

The real-time method \cite{Sekizawa(2022)} offers an intuitive approach
to extract the collective mass of a nuclear cluster immersed in neutron
superfluid. In the latter approach,
we exert an external force $F_\text{ext}$ on protons which are well localized
inside the cluster. The protons together with effectively bound (bound plus
entrained) neutrons start moving towards the direction of the external
force. If the force is constant and there are no redundant excitations
other than the collective translational motion, the cluster would exhibit
a constant acceleration motion, $F_\text{ext}=\dot{P}=M_\text{cluster}a_p$,
where $\dot{P}$ is time derivative of the total linear momentum of the system,
$M_\text{cluster}$ and $a_p$ are the collective mass and the acceleration
of the cluster, respectively. Since protons are localized in space, the center-of-mass
position of protons, $Z_p(t)=\frac{1}{N_p}\int_0^a z n_p(z)\dd z$, is a
well defined quantity. Thus, we can numerically compute the acceleration
of the cluster, $a_p(t)=\dd^2Z_p/\dd t^2$. Adopting the classical relation,
we can evaluate the collective mass of the cluster as well as that of
protons (per unit area) as follows:
\begin{eqnarray}
M_\text{cluster} &=& \frac{\dot{P}_\text{tot}}{a_p},\label{Eq:M_slab}\\
M_p &=& \frac{\dot{P}_p}{a_p},\label{Eq:M_p}
\end{eqnarray}
where $P_\text{tot}=P_n+P_p$ is the total linear momentum and $P_q$
is the linear momentum of neutrons ($q=n$) and protons ($q=p$),
\begin{equation}
P_q(t) = \hbar\int_0^a j_z^{(q)}(z,t)\dd z.
\end{equation}
Since we have both the collective masses of the cluster and of protons, we
can define the collective mass of effectively-bound neutrons per unit area by
\begin{equation}
M_n^\text{eff.bound} = M_\text{cluster}-M_p.
\end{equation}
The number density of conduction neutrons which can freely conduct
can be quantified as follows:
\begin{equation}
n_n^\text{c} = \frac{N_n-N_n^\text{eff.bound}}{a},
\label{Eq:n_conduction}
\end{equation}
where $N_n^\text{eff.bound}=M_n^\text{eff.bound}/m_{n,\text{bg}}^\oplus$
is the average number of the effectively-bound neutrons per unit area,
where $m_{n,\text{bg}}^\oplus=m_n^\oplus[n_n^\text{bg}]$. We note that
the microscopic effective mass becomes equivalent to the bare mass for
$n_n^\text{bg}=0$, and the equation is valid for isolated slabs without
dripped neutrons. Having the conduction neutron number density, $n_n^\text{c}$,
we can quantify the ``macroscopic'' effective mass, $m_n^\star$, as follows:
\begin{equation}
\frac{m_n^\star}{m_{n,\text{bg}}^\oplus} = \frac{n_n^\text{f}}{n_n^\text{c}},
\label{Eq:def_m_star}
\end{equation}
where $n_n^\text{f}$ denotes the (energetically) free neutron number density,
which is the counter part of the energetically-bound neutrons. In this way,
we can extract the macroscopic effective mass from a dynamic response of
the system to the external potential. The real-time method outlined above
can be directly applied also for the superfluid TDDFT, where all complex effects
are automatically included in the description in a fully self-consistent manner.

\begin{figure}[t]
    \centering
    \includegraphics[width=0.9\columnwidth]{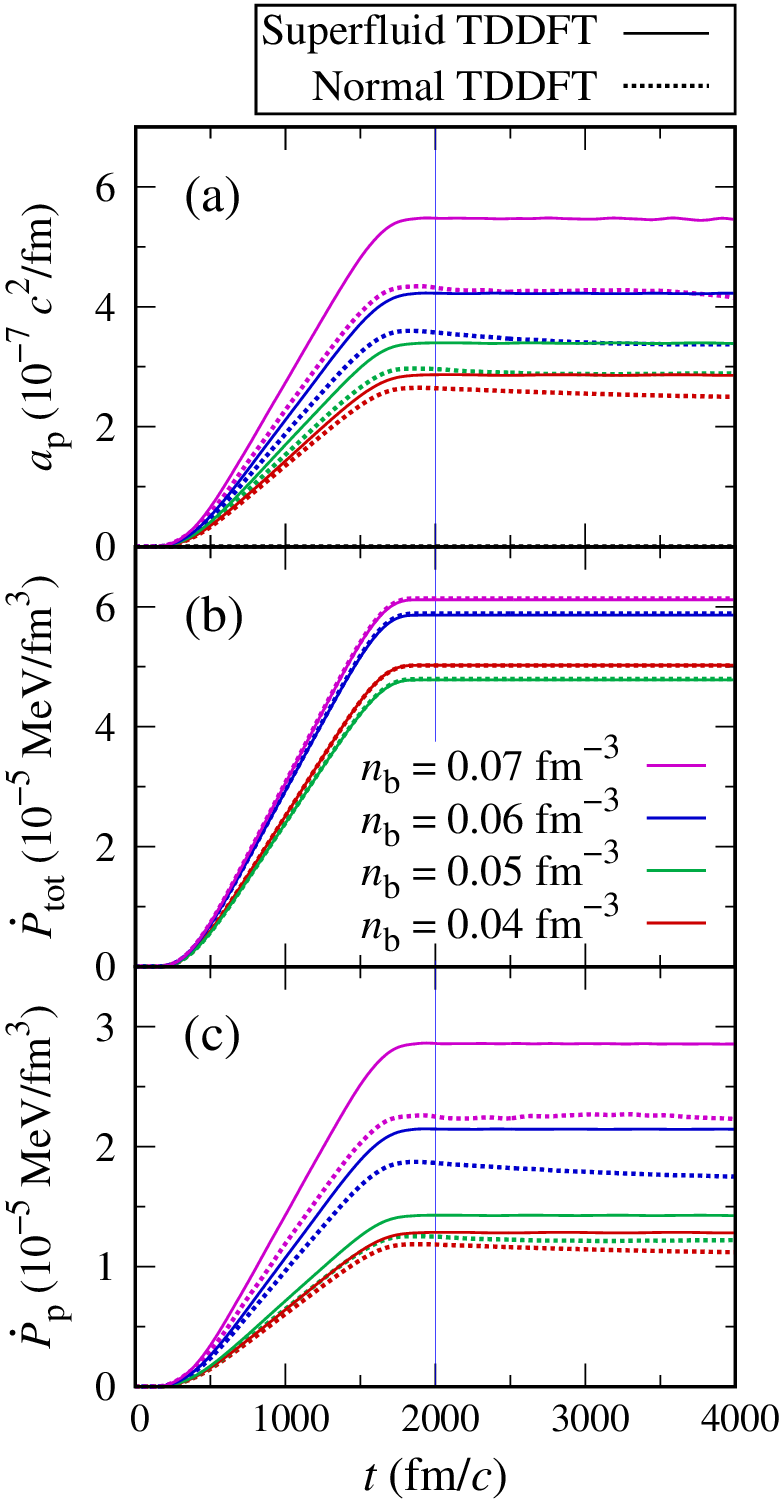}
    \caption{
    Results of fully self-consistent time-dependent band calculations for
    various baryon number densities, $n_\text{b}=0.04$--0.07\,fm$^{-3}$, under the $\beta$-equilibrium
    condition. Results obtained with (without) superfluidity are shown by solid
    (dotted) lines. In panels (a), (b), and (c), the acceleration of the
    center-of-mass position of protons, $a_p$, the rate of change of the total linear
    momentum, $\dot{P}_\text{tot}$, and the rate of change of the proton linear momentum,
    $\dot{P}_p$, are presented, respectively. The vertical line indicates the time
    up to which the external force is turned on.
    }
    \label{accel_nb}
\end{figure}

\subsubsection{The main results}

In Fig.~\ref{accel_nb}, we show obtained acceleration of protons,
$a_p(t)$, in panel (a), time-derivative of the total linear momentum,
$\dot{P}_\text{tot}(t)$, in panel (b), and that of the proton linear
momentum, $\dot{P}_p(t)$, in panel (c), as a function of time. Results
obtained with superfluidity is shown by solid lines, while those obtained
without superfluidity are shown by dotted lines. From $t=0$ to 2,000\,fm/$c$,
we smoothly switch on the external potential and it is kept constant for
$t\ge$\,2,000\,fm/$c$. As a result, we observe that all those quantities
become almost constant for $t\ge$\,2,000\,fm/$c$, as expected. It is to be
mentioned that the acceleration is more stably constant for the superfluid
case (solid lines), as compared with the normal one (dotted lines).
As is shown in Fig.~\ref{accel_nb}(b), the classical relation, $F_\text{ext}
=\dot{P}_\text{tot}$ holds nicely, in both cases with and without superfluidity.
From Figs.~\ref{accel_nb}(a) and (c), we find that both $a_p$ and $\dot{P}_p$
tend to be larger when we include the superfluidity. It indicates that the
collective mass of the slab is lighter when we include superfluidity,
which means that less neutrons are entrained via band structure effects.

\begin{figure}[t]
    \centering
    \includegraphics[width=0.9\columnwidth]{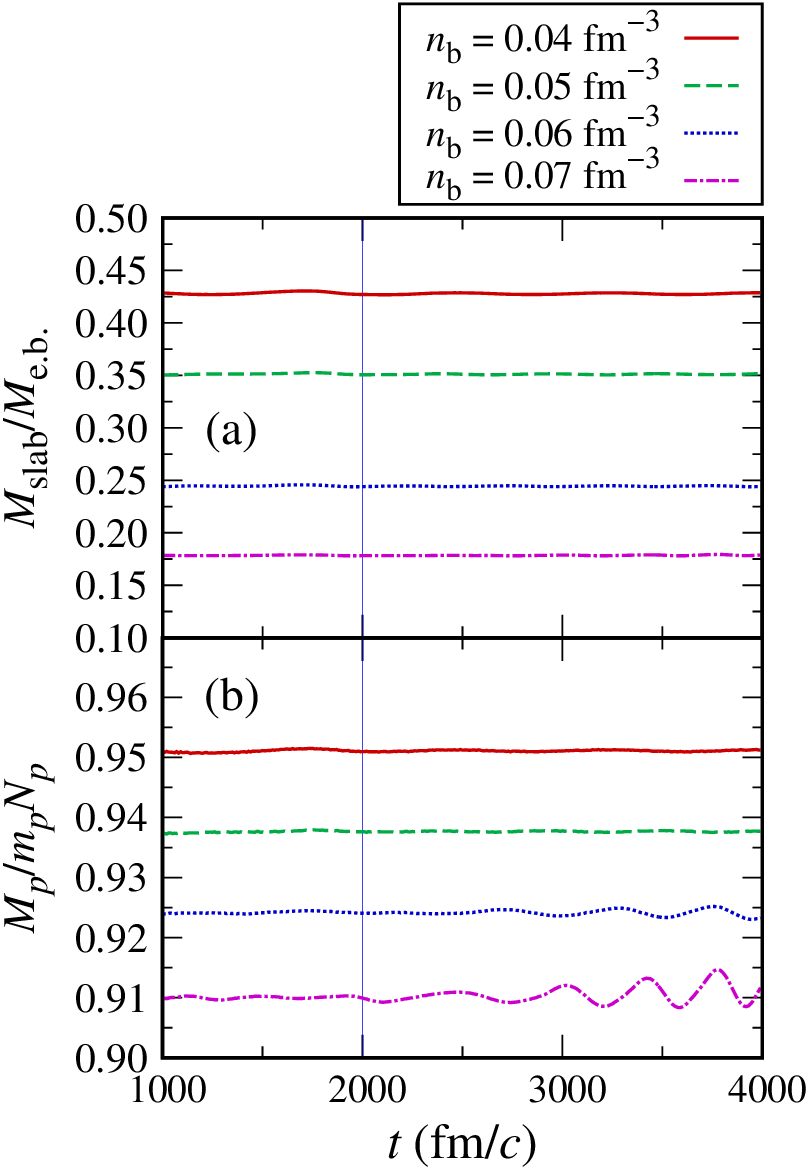}
    \caption{
    Results of fully self-consistent time-dependent superfluid band theory calculations for
    $n_\text{b}=0.04$ (solid line), 0.05 (dashed line), 0.06 (dotted line), and 0.07\,fm$^{-3}$
    (dash-dotted line) under the $\beta$-equilibrium condition. (a) A ratio between the collective
    mass of the slab and the mass of energetically-bound nucleons, $M_\text{slab}/M_\text{e.b.}$,
    is shown as a function of time. (b) A ratio between the collective mass of protons and
    the total mass of protons, $M_p/m_pN_p$, is shown as a function of time. The vertical line
    indicates the time up to which the external force is turned on.
    }
    \label{Mslab}
\end{figure}

Using those quantities presented in Fig.~\ref{accel_nb}, the collective
mass of the slab \eqref{Eq:M_slab} and that of protons \eqref{Eq:M_p}
can be deduced. The results are shown in Fig.~\ref{Mslab}, where the
ratio of $M_\text{slab}$ to the mass of ``energetically-bound'' neutrons
per unit area $M_\text{e.b.}\equiv m_pN_p+m_nN_n^\text{e.b.}$ are shown in panel
(a), and the ratio of $M_p$ to the total mass of protons per unit area
$m_pN_p$ are shown in panel (b). The number of ``energetically-bound''
neutrons, which are within the potential well, is calculated as \cite{Sekizawa(2022)}
\begin{equation}
N_n^\text{e.b.} = \frac{1}{N_{k_z}}\sum_{\nu k_z}\int\frac{k_\parallel}{\pi}
n_{\nu\bm{k}}^{(n)}\theta(U_n^0-e_{\nu\bm{k}}^{(n)})\dd k_\parallel,
\label{Eq:def_N_eb}
\end{equation}
where $\theta(x)$ is the Heaviside step function, where $\theta(x)=0$ for $x<0$
and $\theta(x)=1$ for $x\ge0$. $e_{\nu\bm{k}}^{(n)}$ denotes single-particle energy
after removal of kinetic energy associated with motion parallel to the slabs,
\begin{equation}
e_{\nu\bm{k}}^{(n)} \equiv \varepsilon_{\nu\bm{k}}^{(n)}
-\frac{1}{a}\int_0^a v_{\nu\bm{k}}^{(n)*}(z)\frac{\hbar^2 k_\parallel^2}{2m_n^\oplus(z)}v_{\nu\bm{k}}^{(n)}(z)\dd z.
\end{equation}
From Fig.~\ref{Mslab}(b), we see that the
collective mass of protons is slightly reduced by about 5--9\%. This reduction
was already found in the previous work without superfluidity \cite{Sekizawa(2022)},
which can be explained by the density-dependent microscopic effective mass.
Namely, protons behave as if they have a mass of $m_{p,\text{bg}}^\oplus
\equiv m_p^\oplus[n_{n,\text{b.g.}}]$. In stark contrast, in Fig.~\ref{Mslab}(a),
we find that the collective mass of the slab is significantly \textit{reduced}
by about 57.5--82.5\% from the naive estimation of energetically-bound neutrons
which are within the potential well. Apparently, this significant reduction
can not be explained solely by the density-dependent microscopic effective
mass of neutrons. Note that if dripped neutrons were actually entrained
to the slabs via band structure effects, the collective mass of the slab
would be increased. Therefore, this counter-intuitive phenomenon is
called the ``anti-entrainment'' effects \cite{Sekizawa(2022)}.

\begin{figure}
    \centering
    \includegraphics[width=\columnwidth]{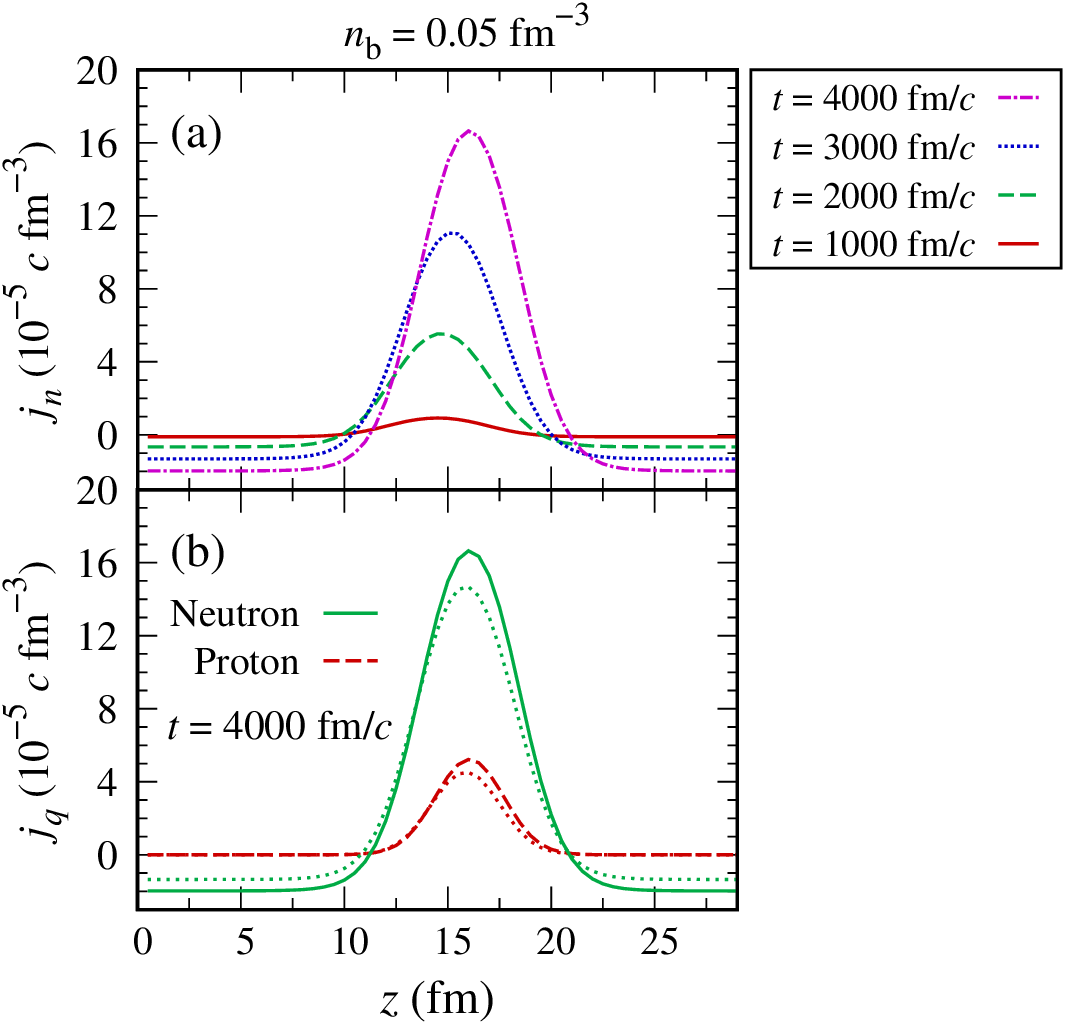}
    \caption{
    Results of fully self-consistent time-dependent superfluid band theory calculations for
    the system with $n_\text{b}=0.05$\,fm$^{-3}$ under the $\beta$-equilibrium condition.
    (a) Neutron current density is shown as a function of the $z$ coordinate at four
    representative instances, $t=\text{1,000}$ (solid line), 2,000 (dashed line),
    3,000 (dotted line), and 4,000\,fm/$c$ (dash-dotted line). (b) Current densities
    of neutrons (solid line) and protons (dashed line) are shown at $t=\text{4,000}$\,fm/$c$.
    Results obtained without superfluidity are also shown in (b) by thin dotted lines,
    for comparison.
    }
    \label{jn_t_nb05}
\end{figure}

As discussed in the previous work \cite{Sekizawa(2022)}, the cause
of the reduction can be found in time evolution of current densities.
In Fig.~\ref{jn_t_nb05}(a), we show neutron current densities obtained
by time-dependent superfluid band theory calculations as a function of $z$
coordinate for the $n_\text{b}=0.05$\,fm$^{-3}$ case. The results are
shown for four representative instances at $t=\text{1,000}$\,fm/$c$ (solid line),
2,000\,fm/$c$ (dashed line), 3,000\,fm/$c$ (dotted line), and 4,000\,fm/$c$
(dash-dotted line). From Fig.~\ref{jn_t_nb05}(a), it is visible that neutrons
around the slab (which was initially $z=a/2=14.5$\,fm) move towards the direction
of the external force ($+z$ direction), because the current density is
positive. On the other hand, the neutron current density outside the slab
becomes negative, meaning that those dripped neutrons move towards $-z$ direction,
opposite to the external force. Since the presence of the ``counterflow''
reduces not only the magnitude of the total linear momentum, $P_\text{tot}$,
but also its rate of increase, $\dot{P}_\text{tot}$, it results in the reduction
of the collective mass of the slab. One could attribute the emergence of the counterflow
to the band structure effects \cite{Sekizawa(2022)}. Namely, since the
macroscopic effective mass depends on the second derivative of single-particle
energy $\varepsilon_{\nu\bm{k}}^{(n)}$ with respect to the Bloch wave number
$k_z$ \cite{Kashiwaba(2019),Sekizawa(2022)}, it can be either positive or
negative depending on the curvature of the band. Dripped neutrons in the band
which is convex upward may have negative macroscopic effective mass, and they
respond towards the opposite direction to the external force. Based on the fully
self-consistent time-dependent superfluid band theory calculations, here we showed
that the anti-entrainment effects do present even with the inclusion of
neutron superfluidity.

To examine the role of superfluidity in the anti-entrainment phenomenon,
we compare the results with and without superfluidity at $t=\text{4,000}$\,fm/$c$
in Fig.~\ref{jn_t_nb05}(b). Normal static and time-dependent band theory calculations
were performed for the same system, keeping all computational parameters unchanged.
Neutron (proton) current density obtained with superfluid band theory is shown by
solid (dashed) line, while those obtained with normal band theory are shown by dotted
lines. From Fig.~\ref{jn_t_nb05}(b), we find that the peak values around the slab
are slightly larger for the superfluid system, consistent with the larger values
of acceleration observed in Fig.~\ref{accel_nb}. In addition, looking at the neutron
current density outside of the slab, we find that the counter flow is also enhanced
with superfluidity. Those observations suggest that the anti-entrainment effect is
slightly enhanced by the inclusion of neutron superfluidity.

\begin{table}[t]
    \centering
    \caption{
    Results of fully self-consistent time-dependent band theory calculations
    with (2nd to 4th columns) and without (5th to 7th columns) superfluidity.
    The baryon number density $n_\text{b}$ is indicated in the first column in
    fm$^{-3}$. In the 2nd and 3rd (5th and 6th) columns, free neutron number
    density $n_n^\text{f}$ and conduction neutron number density $n_n^\text{c}$
    are shown as a ratio to the average neutron number density $\bar{n}_n$,
    respectively. In the 4th (7th) column, macroscopic effective mass $m_n^\star$
    is shown as a ratio to the microscopic effective mass at the background neutron
    number density $m_{n,\text{bg}}^\oplus$, which is calculated with (without) superfluidity.
    }\vspace{2mm}
    \begin{tabular*}{\columnwidth}{@{\extracolsep{\fill}}ccccccc}
        \hline\hline
        & \multicolumn{3}{c}{Superfluid (TD)DFT} & \multicolumn{3}{c}{Normal (TD)DFT}\\ \cline{2-4} \cline{5-7}
        $n_\text{b}$ & $n^\text{f}_n/\bar{n}_n$ & $n^\text{c}_n/\bar{n}_n$ & $m^\star_n/m_{n,\text{bg}}^\oplus$ & $n^\text{f}_n/\bar{n}_n$ & $n^\text{c}_n/\bar{n}_n$ & $m^\star_n/m_{n,\text{bg}}^\oplus$ \\
        \hline
        $0.04$ & $0.702$ & $0.893$ & $0.785$ & $0.710$ & $0.876$ & $0.810$\\ 
        $0.05$ & $0.684$ & $0.913$ & $0.749$ & $0.697$ & $0.896$ & $0.778$\\
        $0.06$ & $0.609$ & $0.933$ & $0.652$ & $0.608$ & $0.911$ & $0.668$\\ 
        $0.07$ & $0.555$ & $0.954$ & $0.582$ & $0.555$ & $0.929$ & $0.598$\\ 
        \hline\hline
    \end{tabular*}
    \label{t_ColEffMass}
\end{table}

From the extracted collective masses shown in Fig.~\ref{Mslab}, we can evaluate
the conduction neutron number density, $n_n^\text{c}$ \eqref{Eq:n_conduction},
and the macroscopic effective mass, $m_n^\star/m_{n,\text{bg}}^\oplus$
\eqref{Eq:def_m_star}. Since the extracted masses slightly fluctuate in time
(cf.\ Fig.~\ref{Mslab}), we take an average over the time interval, 2,000\,fm/$c\le
t\le$\,4,000\,fm/$c$, in which the external force is kept constant. The results
are summarized in Table~\ref{t_ColEffMass}. In the 2nd to 4th columns, results
obtained with superfluidity are shown, while those without superfluidity are
also shown in 5th to 7th columns, for comparison. In both cases with and without
superfluidity, the conduction neutron number density $n_n^\text{c}$ is larger than
the `free' neutron number density $n_n^\text{f}$, for all densities examined,
$n_\text{b}=0.04$--0.07\,fm$^{-3}$. It means that the band structure actually
works like lubricant to mobilize dripped neutrons. As a result, the macroscopic
effective mass $m_n^\star$ becomes smaller than the microscopic effective mass
for background neutron number density $m_{n,\text{bg}}^\oplus$. By comparing
4th and 7th columns, we find that the inclusion of superfluidity slightly
enhances the reduction by a few \%. It is remarkable that the inclusion of
superfluidity does not change the conclusion of Refs.~\cite{Kashiwaba(2019),
Sekizawa(2022)}, at least for the slab phase of the inner crust of neutron stars.

\subsubsection{Remarks on other calculations}

In this section, we discuss relevance and difference of the present results
with other calculations.

First of all, our conclusion of the anti-entrainment in the slab phase looks
contradicting to the results of Carter et~al.~\cite{Carter(2005)}, where
ordinary entrainment effects were reported with $m_n^\star/m_n=1.02$--$1.03$
for $n_\text{b}=0.074$--$0.079$\,fm$^{-3}$ in the slab phase. We consider
that the difference is caused by an improper definition of free neutron
number density, as pointed out in Ref.~\cite{Kashiwaba(2019)}. That is,
in the definition of Carter et~al.~\cite{Carter(2005)}, kinetic energy
associated with the motion parallel to the slab was involved when they
judge if neutrons are below or above the nuclear potential well. Because
of this fact, the free neutron number density, $n_n^\text{f}$, is substantially
overestimated in the calculations of Ref.~\cite{Carter(2005)} that resulted in
entrainment effects with larger values of $m_n^\star/m_n=n_n^\text{f}/n_n^\text{c}$.
We have actually confirmed that $m_n^\star/m_n>1$ is obtained if we use the
definition of Ref.~\cite{Carter(2005)}, in a similar way as reported in
Ref.~\cite{Kashiwaba(2019)}.

Second, in Refs.~\cite{Watanabe(2017),Minami(2022)}, it is advocated that
superfluidity has a strong impact on the entrainment effects and also anti-entrainment
effects are absent, which contradict to our conclusion. We point out here that
it is mainly caused by different theoretical frameworks and definitions of the
effective mass. In Refs.~\cite{Watanabe(2017),Minami(2022)}, a simple toy model
that uses a sinusoidal external potential, which mimics a crystalline structure,
was employed. On top of that external potential, band calculations based on BCS-
and/or HFB-type theories were performed to quantify the macroscopic effective mass
via a standard formula from the band theory of solids. In the simple model of
Refs.~\cite{Watanabe(2017),Minami(2022)}, the effective mass is defined with the
\textit{superfluid fraction}, which is inevitably smaller than or equal to the
total particle number density of the system under the external potential. It is
thus by definition impossible to obtain the anti-entrainment effect in their model.
On the contrary in our framework, the effective mass is defined as the ratio of
the number density of \textit{energetically free} neutrons, which are seemingly
free from the potential, to the number density of conduction neutrons, which can
actually conduct. We note that our fully self-consistent calculations showed that
more than $90$\% of neutrons actually participate in conduction (cf.\ the 3rd and
6th columns of Table~\ref{t_ColEffMass}). It underlines the importance of the self-organizing
character of nuclear systems, where neutrons and protons form clusters by themselves
through the nucleon-nucleon interaction, and there is no external periodic potential.
The latter property is in stark contrast to ordinary solids, where electrons are assumed,
to a good approximation, to feel an external periodic potential of an ionic lattice
in the sense of the Born-Oppenheimer approximation.

Lastly, it is to mention that we have also examined two static treatments for calculating
the neutron effective mass. The first one is the method that was employed in, e.g.,
Refs.~\cite{Carter(2005),Chamel(2005),Chamel(2012),Kashiwaba(2019),Sekizawa(2022)}.
In the latter method, the effective mass is evaluated with the mobility coefficient
which is calculated as a sum of inverse effective mass tensors which depend on curvature
of each band as a function of the Bloch wave number. The other one is the method that
was employed in Ref.~\cite{Magierski(2021)}. In the latter case, TDSLDA equations are
solved in a gauge that simulates induced superflow, and the effective mass of an impurity
can be extracted from a static response of the system. With both approaches, we have
obtained the results which are qualitatively consistent with our real-time method,
indicating the anti-entrainment effects.

Since there remains some room for discussion, such as the definition of free neutrons
and the relationship between the superfluid fraction and the conduction neutron number
density, we leave quantitative comparisons between those different treatments as a future work.

\subsection{Quasiparticle Resonances}\label{Sec:qpresonances}

Finally, here we report our complementary finding of an intriguing phenomenon
characteristic to superfluid systems, known as \textit{quasiparticle resonances},
in the context of the inner crust of neutron stars and discuss if it is relevant
to the entrainment phenomenon.

Usually, quasiparticle resonances are studied in the context of a usual neutron scattering process
in a vacuum \cite{Dobaczewski(1996),Bennaceur(1999),Bulgac(2000),Kobayashi(2016),
Kobayashi(2020)}. It is a resonance associated with pairing correlations,
where an incoming neutron deposits part of its kinetic energy to the target nucleus,
inducing a particle-hole excitation of a bound neutron, and the excited neutron
and the incoming one form a Cooper pair which behaves as a resonance.
Here we show, within the fully self-consistent superfluid band theory calculations,
that quasiparticle resonances do present even in the inner crust of neutron stars,
where superfluid neutrons permeate crystalline nuclear matter.

An intuitive way to distinguish quasiparticle resonances among the others
is to analyze the occupation probabilities as a function of single-particle
energy. However, what we obtain as a solution of the SLDA equation is a set
of quasiparticle energies, not the single-particle ones, and one should
evaluate the latter in an appropriate manner. In the present work, instead
of introducing the canonical basis, we use an alternative method explained below.

As an effective way to link quasiparticle energies with single-particle ones,
we take the following procedure:
\begin{enumerate}
\item We solve the SLDA equation self-consistently and obtain quasiparticle
energies $E_{\nu\bm{k}}^{(q)}$ as well as a set of densities, say, $\bm{\rho}_\text{g.s.}$,
and the corresponding single-particle Hamiltonian $h_{\bm{k},\mathrm{g.s.}}^{(q)}\equiv
h^{(q)}[\bm{\rho}_\text{g.s.}]+h_{\bm{k}}^{(q)}[\bm{\rho}_\text{g.s.}]$.

\item Next, we diagonalize the single-particle Hamiltonian $h_{\bm{k},\mathrm{g.s.}}^{(q)}$
and obtain `effective' single-particle energies, say, $\breve{\varepsilon}_{\nu'\bm{k}}^{(q)}$.
(Note that we put a breve accent to indicate that they are not necessarily exactly equal
to the true single-particle energies, $\varepsilon_{\nu'\bm{k}}^{(q)}$.) The indices
$\{\nu'\bm{k}\}$ are labeling the effective single-particle energies, $\breve{\varepsilon}
_{\nu'\bm{k}}^{(q)}$, in the ascending order.

\item We then evaluate the corresponding quasiparticle energies using the relation,
\begin{equation}
\breve{E}_{\nu'\bm{k}}^{(q)}
= \sqrt{(\breve{\varepsilon}_{\nu'\bm{k}}^{(q)}-\lambda_q)^2+\overline{\Delta}_q^2},
\label{Eq:breve_qpenergies}
\end{equation}
where $\overline{\Delta}_q$ represents the average absolute value of the pairing field,
defined in Eq.~\eqref{Eq:average_Delta_q}.

\item We reorder the obtained quasiparticle energies, $\breve{E}_{\nu'\bm{k}}^{(q)}$
\eqref{Eq:breve_qpenergies}, in the ascending order, $\breve{E}_{\nu'\bm{k}}^{(q)}
\rightarrow\breve{E}_{\nu\bm{k}}^{(q)}$, where we store the correspondence between
the indices, $\{\nu'\bm{k}\}\Leftrightarrow\{\nu\bm{k}\}$.

\item Based on the following correspondences,
\begin{equation}
\varepsilon_{\nu'\bm{k}}^{(q)}
\;\approx\;
\breve{\varepsilon}_{\nu'\bm{k}}^{(q)}
\,\Leftrightarrow\,
\breve{E}_{\nu\bm{k}}^{(q)}
\;\approx\;
E_{\nu\bm{k}}^{(q)},
\end{equation}
we regard that the single-particle energies $\varepsilon_{\nu'\bm{k}}^{(q)}$
are associated with the corresponding states $\{\nu'\bm{k}\}\Leftrightarrow
\{\nu\bm{k}\}$ having occupation probabilities,
\begin{equation}
n_{\nu\bm{k}}^{(q)} = \frac{1}{a}\int_0^a \bigl|v_{\nu\bm{k}}^{(q)}(z)\bigr|^2 \dd z.
\end{equation}
\end{enumerate}
The estimated quasiparticle energies, $\breve{E}_{\nu\bm{k}}^{(q)}$
\eqref{Eq:breve_qpenergies}, may not exactly be equal to the true ones,
$E_{\nu\bm{k}}^{(q)}$, but in the most cases we found a good correspondence,
$\breve{E}_{\nu\bm{k}}^{(q)}\simeq E_{\nu\bm{k}}^{(q)}$, with a correct
ordering of the quasiparticle energies.

\begin{figure}
    \centering
    \includegraphics[width=0.95\columnwidth]{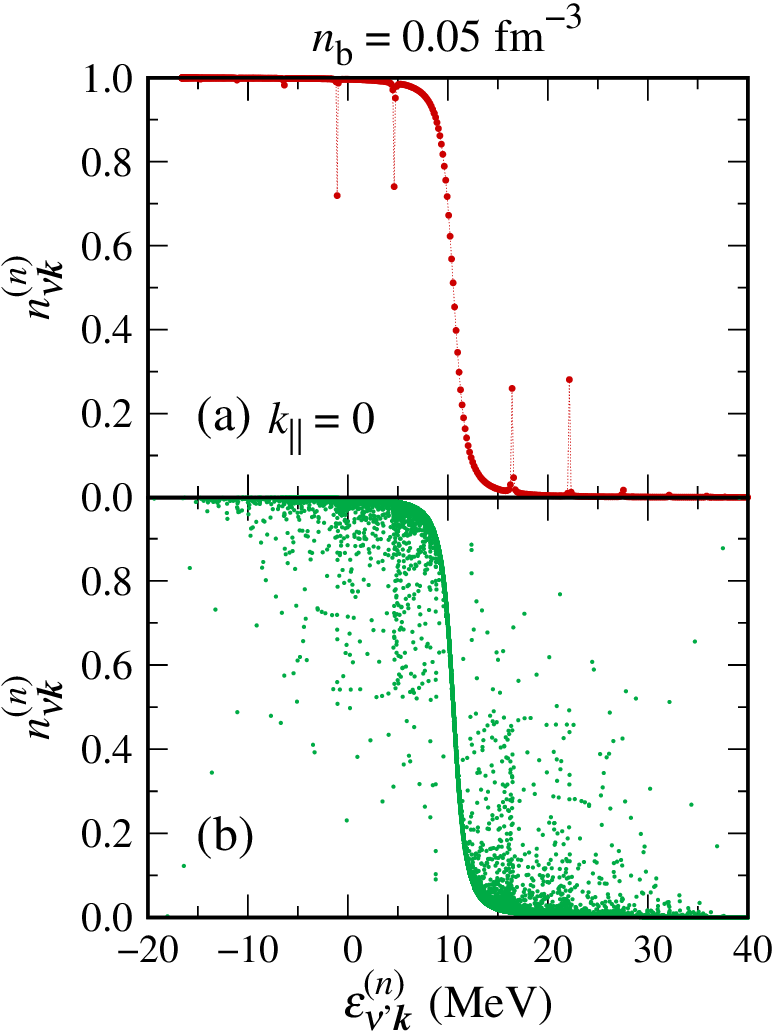}
    \caption{
    Occupation probabilities of neutrons, $n_{\nu\bm{k}}^{(n)}$, are shown as
    a function of single-particle energy, $\varepsilon_{\nu'\bm{k}}^{(n)}$,
    for states with $0\leq k_z \leq \pi/a$ and $n_\text{b}=0.05$\,fm$^{-3}$
    under the $\beta$-equilibrium condition. In panel (a), occupation probabilities
    are plotted only for states with $k_\parallel=0$ for better visibility, while
    in panel (b), results are plotted for all states.
    }
    \label{Energy_occ_nb05}
\end{figure}

In Fig.~\ref{Energy_occ_nb05}, we show neutron occupation probabilities
$n_{\nu\bm{k}}^{(n)}$ as a function of single-particle energies
$\varepsilon_{\nu'\bm{k}}^{(n)}$, calculated for the system with $n_\text{b}=0.05$\,fm$^{-3}$
under the $\beta$-equilibrium condition, as a typical example. Since single-particle energies
are the same for $\pm k_z$, results are plotted for a half of the first Brillouin
zone, $0\le k_z\le \pi/a$. The plot is restricted to $k_\parallel=0$ in
Fig.~\ref{Energy_occ_nb05}(a) for better visibility, while occupation probabilities
for all states within the plotting range, $-20\,\text{MeV}\le \varepsilon
_{\nu'\bm{k}}^{(n)}\le40\,\text{MeV}$, are presented in Fig.~\ref{Energy_occ_nb05}(b).

First, let us focus on the $k_\parallel=0$ case, shown in Fig.~\ref{Energy_occ_nb05}(a).
From the figure, we see that occupation probabilities globally follow a Fermi-Dirac-type
distribution, as it should be, indicating that the above mentioned procedure works
well. In addition to that, we find that there appear several irregular dips
and peaks below and above the chemical potential $\lambda_n\simeq10.8$\,MeV, respectively.
Clearly, the peaks embedded in the continuum, paired up with the dips in the bound
states, manifest the expected characteristic of quasiparticle resonances,
that is, they form Cooper pairs to gain pairing energy.

\begin{figure}[t]
    \centering
    \includegraphics[width=\columnwidth]{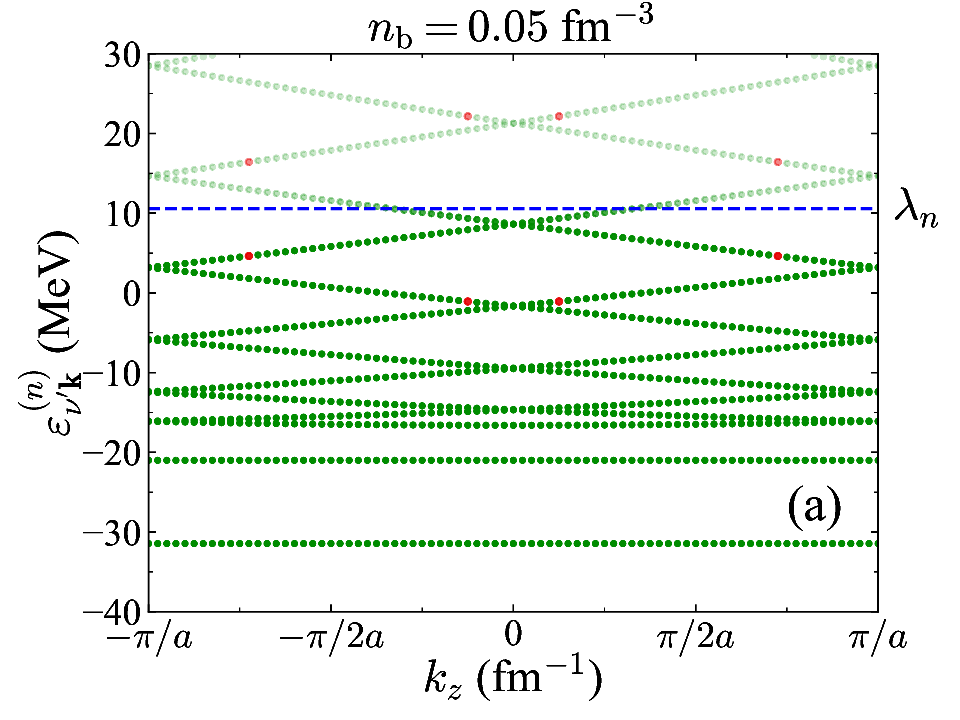}\vspace{4mm}
    \includegraphics[width=\columnwidth]{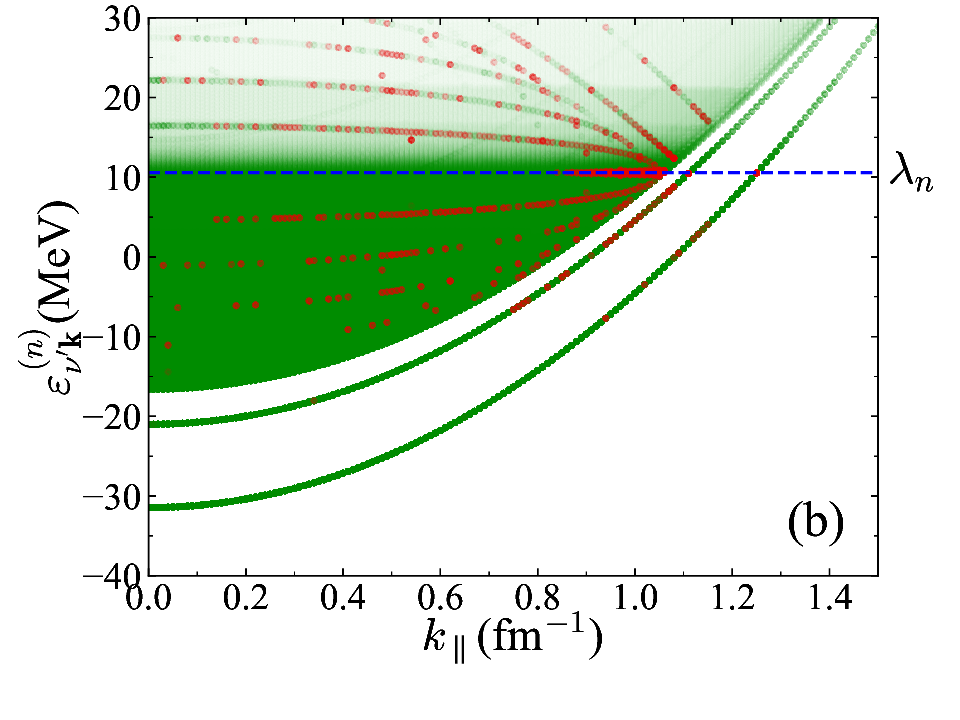}\vspace{-6mm}
    \caption{
    Single-particle energies of neutrons, $\varepsilon_{\nu'\bm{k}}^{(n)}$,
    are shown as functions of $k_z$ for $k_\parallel=0$ (a) and as functions
    of $k_\parallel$ for $0\le k_z\le\pi/a$ (b), for the system with $n_\text{b}=
    0.05$\,fm$^{-3}$ under the $\beta$-equilibrium condition. Horizontal dashed
    line indicates the neutron chemical potential $\lambda_n$. Occupation probabilities
    $n_{\nu\bm{k}}^{(n)}$ are represented by opacity of symbols. In both panels,
    states regarded as quasiparticle resonances are highlighted in red color.
    }
    \label{res_nb05}
\end{figure}

To serve another look at its behavior, we show in Fig.~\ref{res_nb05}(a)
single-particle energies of neutrons $\varepsilon_{\nu'\bm{k}}^{(n)}$ as a
function of the Bloch wave number $k_z$ in the first Brillouin zone, $-\pi/a
\le k_z\le\pi/a$, with $k_\parallel=0$. In the figure, occupation probabilities
are indicated by the opacity of the data symbols. From the figure, we find that
the single-particle energies nicely exhibit the expected band structure (cf.\
Refs.~\cite{Kashiwaba(2019),Sekizawa(2022)}). On top of that, we can clearly
see that there are four irregular states above the chemical potential, which
sustain noticeable occupation probabilities. A closer look at the results reveals
that hole-like states with relatively small occupation probabilities are present
at the same $k_z$ below the chemical potential, meaning that Cooper pairs are
formed between the states with the same Bloch wave vector $\bm{k}$. We note that
these quasiparticle resonances could be observed, thanks to the band theory
calculations, i.e., they are absent for the $k_z=0$ case, in this particular
example. We consider that the essence is a better treatment of the continuum
states---the resonances would not be resolved if there were no pairs of states
with close values of $|\varepsilon_{\nu'\bm{k}}^{(n)}-\lambda_n|$ within
a resonance width. In addition, we plot in Fig.~\ref{res_nb05}(b) single-particle
energies as a function of $k_\parallel$ as well. The states regarded as quasiparticle
resonances are highlighted by red color. From these figures, we find that pairing
is simply associated with single-particle states around the chemical potential.
If there were a large band gap ($\Delta\varepsilon>\overline{\Delta}_n$)
and the chemical potential were located in between the gap, pairing would have
been suppressed. In the slab phase, however, band gaps are only a few to tens
of keV \cite{Kashiwaba(2019),Sekizawa(2022)}, which are much smaller than the
average absolute value of the pairing field, $\overline{\Delta}_n\approx1$\,MeV.
Therefore, pairing properties are insensitive to the band structure, in the
slab phase under study.

In Fig.~\ref{Energy_occ_nb05}(b), we show occupation probabilities for all states
($k_z>0$ and $k_\parallel\ge0$) as a function of single particle energies. Intriguingly,
we find that there are a number of quasiparticle resonances for a range of single-particle
energies, as can be seen from the figure. Thus, on top of the usual entrainment
effects, where one expects that part of dripped neutrons are effectively immobilized
by the periodic potential, there could be an additional contribution from quasiparticle
resonances which are detectable only in microscopic superfluid calculations.

To estimate the impact of quasiparticle resonances on the entrainment phenomenon,
we calculate the total number of resonating neutrons per unit area, $N_n^\text{res}$.
$N_n^\text{res}$ is calculated by integrating densities of states above the
chemical potential, but with relatively large occupation probabilities greater
than 0.1, that we considered as candidates of the quasiparticle resonances. To exclude the
states having $n_{\nu\bm{k}}^{(n)}>0.1$ in the tail of the global Fermi-Dirac-type
distribution, only those exhibiting sudden changes of occupation probabilities
in neighboring energies are regarded as quasiparticle resonances. The results
for various densities, $n_\text{b}=0.04$, 0.05, 0.06, and 0.07\,fm$^{-3}$,
are summarized in Table.~\ref{Tab_resonance}. In the 5th column of Table~\ref{Tab_resonance},
we show the total number of resonating neutrons per unit area within a single slab
period $a$, $N_n^\text{res}$. These numbers should be compared with the expected
number of bound neutrons. Here we take a ratio between the number of resonating
neutrons to the number of ``energetically-bound'' neutrons \eqref{Eq:def_N_eb},
$N_n^\text{res}/N_n^\text{e.b.}$, and it is listed in the 6th column of Table~\ref{Tab_resonance}.
From the results, we find that the number of resonating neutrons is only less
than or around 1\% of the number of energetically-bound neutrons in the system.
We also find that the number of resonating neutrons decreases as the baryon
number density increases. The latter observation can be explained by looking
at the change of the potential. In the 2nd, 3rd, and 4th columns of Table~\ref{Tab_resonance},
the minimum and the maximum values of the mean-field potential, $U_n^\text{min}$
and $U_n^0$, respectively, and the chemical potential, $\lambda_n$,
are presented. Because more and more neutrons are dripped out of the slabs
as the baryon number density increases, the maximum value of the mean-field
potential decreases substantially, while the minimum value is not so much
affected. It means that the depth of the potential well becomes shallower
for higher baryon number densities and, as a result, there are less bound
orbitals which can contribute to the quasiparticle resonances. Therefore,
although the phenomenon itself is physically intriguing, its impact on
the entrainment effect is negligibly small.

\begin{table}[t]
    \centering
    \caption{
    Results of the SLDA calculations for baryon number densities $n_\text{b}=
    0.04$, 0.05, 0.06, and 0.07\,fm$^{-3}$ under the $\beta$ equilibrium condition
    are listed in 2nd--5th rows, respectively. From left to right columns, it shows:
    baryon number density, $n_\text{b}$, in fm$^{-3}$, the minimum and the maximum
    values of the mean-field potential, $U_n^\text{min}$ and $U_n^0$, in MeV, the
    neutron chemical potential, $\lambda_n$, in MeV, the number of resonating
    neutrons per unit area, $N_n^\text{res}$, in fm$^{-2}$, and its ratio to the
    number of energetically-bound neutrons, $N_n^\text{res}/N_n^\text{e.b}$
    [cf.\ Eq.~\eqref{Eq:def_N_eb}].
    }\vspace{3mm}
    \begin{tabular*}{\columnwidth}{@{\extracolsep{\fill}}cccccc}
        \hline\hline$n_\text{b}$ & $U_n^{\mathrm{min}}$ & $U_n^0$ & $\lambda_n$ & $N_n^\text{res}$ & $N_n^\text{res}/N_n^\text{e.b.}$ \\
        \hline
        $0.04$ & $-37.3$ & $-13.1$ & $9.48$ & $1.09\times 10^{-2}$ & 2.49\% \\ 
        $0.05$ & $-37.6$ & $-16.7$ & $10.6$ & $3.92\times 10^{-3}$ & 0.88\% \\ 
        $0.06$ & $-39.4$ & $-20.3$ & $11.6$ & $2.54\times 10^{-3}$ & 0.36\% \\ 
        $0.07$ & $-38.7$ & $-23.8$ & $12.7$ & $1.05\times 10^{-3}$ & 0.14\% \\
        \hline\hline
    \end{tabular*}
    \label{Tab_resonance}
\end{table}

\section{Conclusion and Prospect}\label{Sec:Conclusion}

In this work, we have developed a fully self-consistent time-dependent superfluid
band theory for the inner crust of neutron stars, based on time-dependent density
functional theory (TDDFT) extended for superfluid systems, known as time-dependent
superfluid local density approximation (TDSLDA). It should be noted that our theoretical
framework, though it is yet restricted to 1D slab structure, is much more realistic
than other existing models on the market. Namely, our theory is based on the microscopic
framework of (TD)DFT, which can correctly describe properties of finite nuclei, not
only static structure, but also excitations and reaction dynamics, as well as nuclear
matter properties, in a unified way. We do not \textit{a priori} assume any external
potential nor cluster shape, whereas nuclear pasta is formed self-consistently through
the effective nucleon-nucleon interaction. By applying the real-time method, where
we measure the collective mass of a nuclear cluster immersed in neutron superfluid
through a response of the system to an external force, we have successfully extracted
the conduction neutron number density and the macroscopic effective mass of dripped
neutrons. From the results, we have found that the dripped neutrons are actually
mobilized by the band structure, that is, the conduction neutron number density is
enhanced and the neutron effective mass is reduced, which we call the \textit{anti-entrainment}
effects. These results are consistent with recent self-consistent band calculations
without superfluidity \cite{Kashiwaba(2019),Sekizawa(2022)}, that is, no significant
qualitative difference was observed in the cases with and without superfluidity.
We have demonstrated that the neutron effective mass is substantially reduced
up to about 42\% in the slab phase and superfluidity slightly enhances this
anti-entrainment effect. 

As a next step, we have already extended the present formalism to include finite
temperature and magnetic field effects (cf.\ \cite{Sekizawa(2023)}). It enables us
to quantify, e.g., the melting temperature of the slabs, taking into account the band
structure effects. We expect that such fully self-consistent finite-temperature band
theory calculations of nuclear pasta phases in a hot environment will be useful, e.g.,
for studying supernova matter or cooling of proto-neutron stars. Interesting and useful
information such as neutrino-pasta scattering, elastic properties, as well as neutron-star
cooling will be obtained in our forthcoming works.

Needless to say, it is highly desired to extend the present work to 2D and 3D geometries.
We believe that the formalism itself is unchanged and can be extended to higher dimensions
in a straightforward way. Thus, the major obstacle is the computational cost. As we
mentioned in Sec.~\ref{Sec:ComptDetails}, we have already dealt with millions of
quasiparticle wave functions for the 1D geometry. The extensions to 2D and 3D geometries
would require tens to thousands times larger number of quasiparticle orbitals and the
number of numerical operations for those lattice points would be increased. To avoid
diagonalizations of a matrix with such a huge dimension, we may take advantage of shifted
conjugate-orthogonal conjugate-gradient (COCG) \cite{Jin(2017)} or shifted conjugate-orthogonal
conjugate-residual (COCR) \cite{Kashiwaba(2020)} methods. These methods allow us to extract
various densities by contour integrations, where a shifted algorithm can be used to efficiently
evaluate quantities at different points in the complex plane. It has been shown that these methods
are particularly suitable to GPU parallelization, since a huge number of operations at
different coordinates $\bm{r}$ are independent from each other. In addition, TDSLDA has
also been shown remarkable successes with the use of top-tier supercomputers working with
GPUs \cite{Jin(2021)}. The use of GPUs would enable us to realize fully self-consistent
time-dependent superfluid band theory calculations for 2D and 3D geometries and to resolve
the controversial situation concerning the entrainment effects in the inner crust
of neutron stars.

\begin{acknowledgments}
We are grateful to Takashi Nakatsukasa (University of Tsukuba) and
Kenichi Yoshida (RCNP, Osaka University) for valuable discussions.
We also thank Giorgio Almirante (IJCLab) for communication that resolved
an error in our computational code.
Meetings in the A3 Foresight Program supported by JSPS are also acknowledged
for useful discussions. One of the authors (K.Y.) would like to acknowledge
the support from the Hiki Foundation, Tokyo Institute of Technology.
This work mainly used computational
resources of the Yukawa-21 supercomputer at Yukawa Institute for Theoretical
Physics (YITP), Kyoto University. This work also used (in part) computational
resources of the HPCI system (Grand Chariot) provided by Information Initiative
Center (IIC), Hokkaido University, through the HPCI System Project (Project ID:
hp230180). This work is supported by JSPS Grant-in-Aid for Scientific Research,
Grant Nos.~23K03410 and 23K25864.
\end{acknowledgments}

\appendix

\section{On total energy per nucleon}\label{App:Etot/A}

\begin{figure*}
    \centering
    \includegraphics[width=\textwidth]{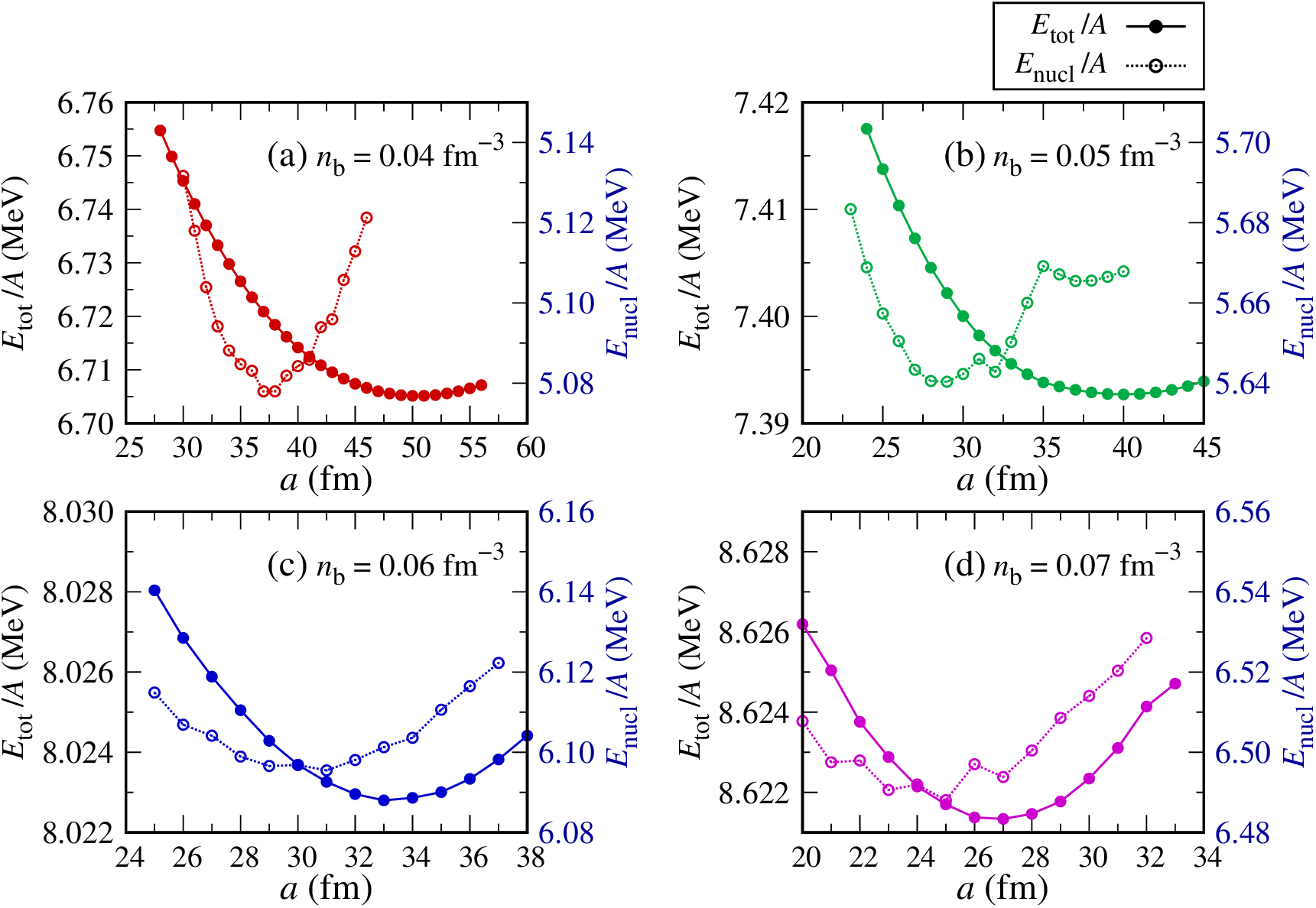}
    \caption{
    Energy per nucleon is shown as a function of the slab period $a$.
    Total energy per nucleon which include electrons' contribution, $E_\text{tot}/A$,
    is shown by solid circles connected with solid lines, while that without
    electrons contribution, $E_\text{nucl}/A$, is shown by open circles connected
    with dotted lines. In panels (a)--(d), results obtained for different baryon
    number densities, $n_\text{b}=0.04$, $0.05$, $0.06$, and $0.07$\,fm$^{-3}$,
    respectively, under the $\beta$-equilibrium condition are presented. Note that
    left and right vertical axes correspond to $E_\text{tot}/A$ and $E_\text{nucl}/A$,
    respectively.
    }
    \label{Fig:Etot/A}
\end{figure*}

As we mentioned in Sec.~\ref{Sec:a_opt}, slab-period dependence of energy
of the system shows different behavior with and without electrons' contribution.
In this Appendix, we provide a comparison between those two cases.

In Figs.~\ref{Fig:Etot/A}(a)--(d), we present the total energy per nucleon
for $n_\text{b}=0.04$, $0.05$, $0.06$, and $0.07$\,fm$^{-3}$, respectively,
under the $\beta$-equilibrium condition as a function of the slab period $a$.
The results with electrons' contribution ($E_\text{tot}/A$) are shown by filled
circles connected with solid lines, while those without electrons' contribution
($E_\text{nucl}/A$) are shown by open circles connected with dotted lines.
That is, the latter results are the same as those shown in Fig.~\ref{AvsE_nb05}
in the main texts.

From the figure, we find that both energies, $E_\text{tot}/A$ and $E_\text{nucl}/A$,
exhibit a parabolic shape which is convex downward as a function of the slab period $a$.
The minimum energy locations are shifted towards larger $a$ values by
12, 11, 2, 2\,fm for $n_\text{b}=0.04$, $0.05$, $0.06$, and $0.07$\,fm$^{-3}$.
We find that the shift is smaller for higher density region where emergence of
the slab phase is actually expected ($n_\text{b}\simeq0.07$--$0.08$\,fm$^{-3}$).
To obtain an equilibrium configuration of nuclear pasta phases under the
$\beta$-equilibrium condition, the total energy, rather than the nuclear energy,
should be minimized. Because the use of larger $a$ values (i.e., larger $N_z$)
requires substantial computational effort, and we expect that it will not change
the conclusion of the present article, the configurations that minimize the
nuclear energies were analyzed in this work.

\end{document}